\documentclass[12pt]{article}
\usepackage{amsmath}
\usepackage[dvips]{graphicx}        %pacchetto per inserire immagini
\usepackage{subfigure}
\usepackage{textcomp}
\input{epsf} 
\def\ltap{\raisebox{-.6ex}{\rlap{$\,\sim\,$}} \raisebox{.4ex}{$\,<\,$}}

\newcommand\as{\alpha_{\mathrm{S}}} 
\newcommand\f[2]{\frac{#1}{#2}} 
\def\beq{\begin{equation}} 
\def\eeq{\end{equation}} 
\def\beeq{\begin{eqnarray}} 
\def\eeeq{\end{eqnarray}} 
 
\def\to{\rightarrow}
 
\def\nn{\nonumber} 
 
\def\ptmin{p_{T{\rm min}}}
\def\ptmax{p_{T{\rm max}}}

\def\tL{{\widetilde L}}

\begin{document} 

\begin{titlepage}
\begin{flushright}
ZU-TH 04/12\\
IFUM-993-FT
\end{flushright}

\renewcommand{\thefootnote}{\fnsymbol{footnote}}
\vspace*{1.5cm}

\begin{center}
{\Large \bf  Higgs boson production at the LHC:\\[0.2cm] transverse momentum resummation effects in the\\[0.4cm]$H\to\gamma\gamma$, $H\to WW\to l\nu l\nu$ and $H\to ZZ\to 4l$ decay modes}
\end{center}
\par \vspace{2mm}
\par \vspace{2mm}
\begin{center}
\vspace{.5cm}
{\bf Daniel de Florian}$^{(a,b)}$, 
{\bf Giancarlo Ferrera}$^{(b,c)}$,\\ {\bf Massimiliano Grazzini}$^{(b)}$\footnote{On leave of absence from INFN, Sezione di Firenze, Sesto Fiorentino, Florence, Italy.}~~and~~{\bf Damiano Tommasini}$^{(b,d)}$\\

\vspace{0.8cm}

${}^{(a)}$
Departamento de F\'\i sica, FCEYN, Universidad de Buenos Aires,\\
(1428) Pabell\'on 1 Ciudad Universitaria, Capital Federal, 
Argentina\\\vskip .05cm
${}^{(b)}$Institut f\"ur Theoretische Physik, Universit\"at Z\"urich, CH-8057 
Z\"urich, Switzerland\\\vskip .05cm
${}^{(c)}$
Dipartimento di Fisica, Universit\`a di Milano and\\
INFN, Sezione di Milano, I-20133 Milan, Italy\\
${}^{(d)}$
Dipartimento di Fisica e Astronomia, Universit\`a di Firenze and\\
INFN, Sezione di Firenze,
I-50019 Sesto Fiorentino, Florence, Italy
\vspace{1cm}
\end{center}

\vspace{5mm}

\par \vspace{2mm}
\begin{center} {\large \bf Abstract} \end{center}
\begin{quote}
\pretolerance 10000

We consider Standard Model Higgs boson production through gluon--gluon
fusion in hadron collisions. We combine the calculation of the next-to-next-to-leading order QCD corrections to the inclusive cross section
with the resummation of multiple
soft-gluon emissions at small transverse momenta up to next-to-next-to-leading logarithmic accuracy.
The calculation is implemented in the numerical program {\tt HRes} and
allows us to retain the full kinematics of the Higgs boson and of its decay products.
We present selected numerical results for the signal cross section
at the LHC ($\sqrt{s}=8$ TeV), in the $H\to\gamma\gamma$, $H\to WW\to l\nu l\nu$ and $H\to ZZ\to 4l$ decay channels by using the nominal cuts applied in current Higgs boson searches by the ATLAS and CMS collaborations.
\end{quote}

\vspace*{\fill}
\begin{flushleft}
%\date
March 2012

\end{flushleft}
\end{titlepage}

\setcounter{footnote}{1}
\renewcommand{\thefootnote}{\fnsymbol{footnote}}

\section{Introduction}

One of the main tasks of the LHC program is the search for the Higgs boson \cite{Englert:1964et}
and the study of its properties (mass, couplings,
decay widths).

The LHC, after a successful start of $pp$ collisions in 2009 and 2010, has been operated at a centre-of-mass energy of $7$ TeV in 2011, and data
corresponding to an integrated luminosity of 5.7 fb$^{-1}$ have been accumulated.
These data already allowed the ATLAS \cite{atlas} and CMS \cite{cms} experiments to shrink the
allowed mass range for the Standard Model (SM) Higgs boson $H$ considerably by essentially excluding
the Higgs bosons in the mass range ${\cal O}$(130 GeV) $< m_H < {\cal O}$(600 GeV), while observing an excess of Higgs boson candidate events around $m_H= 125$ GeV. An update of the Tevatron results \cite{tevatron} with up
to $10$ fb$^{-1}$ integrated luminosity shows a broad excess of events in the
region $115-135$ GeV. 
More data from the LHC 2012 run, that will be operated at a  centre-of-mass energy of $8$ TeV, are
needed to say whether these excesses really correspond to a Higgs signal or are just statistical fluctuations.

In this paper we consider the production 
of the SM Higgs boson by the gluon fusion mechanism and its decays
$H\to\gamma\gamma$, $H\to WW$ and $H\to ZZ$.
The gluon fusion process $gg \to H$ \cite{Georgi:1977gs}, through a heavy-quark loop,
is the main production mechanism of the 
SM Higgs boson at hadron colliders. 
The corresponding cross section is typically at least
one order of magnitude larger
than the cross section in the other production channels (vector boson fusion, associated production....), and becomes comparable with the cross section
for vector boson fusion
only at high Higgs boson masses.
It is thus essential to achieve reliable theoretical
predictions for the gluon fusion cross section
and the associated distributions.
The dynamics of the gluon fusion mechanism is driven by 
strong interactions. Thus, accurate studies of the effect of QCD radiative 
corrections are mandatory 
to obtain precise theoretical predictions.

The leading order (LO) cross section is proportional
to $\as^2$, $\as$ being the QCD coupling.
The QCD radiative corrections to the total cross section have been computed at the
next-to-leading order (NLO)
in Refs.~\cite{Dawson:1990zj,Djouadi:1991tka,Spira:1995rr}
and found to be of the same order as the LO contribution.
The next-to-next-to-leading order (NNLO) corrections have been computed in Refs.
\cite{Harlander:2002wh,Anastasiou:2002yz,Ravindran:2003um}
and their effect is moderate: for a light Higgs,
they increases the NLO cross result by about $25\%$ at the LHC ($\sqrt{s}=8$ TeV).
We recall that all the NNLO results have been obtained by using 
the large-$M_t$ approximation, $M_t$ being the mass of the top quark.
Corrections beyond this approximation have been considered in Refs.~\cite{Marzani:2008az,Harlander:2009bw,Harlander:2009mq,Harlander:2009my,Pak:2009bx,Pak:2009dg}.

The NNLO result mentioned above is certainly important, but
it refers to a fully inclusive cross section.
The impact of higher-order corrections
generally depends on the selection cuts used in the
experimental analysis and also the shape of the
distributions is typically affected by the applied cuts.

A first step in the direction of taking selection cuts into account
was taken in Ref.~\cite{Catani:2001cr},
where the inclusive cross section with a jet veto was computed at NNLO.
The first NNLO calculation of the Higgs production cross section
that fully takes into account experimental cuts
was reported in Ref.~\cite{Anastasiou:2005qj}, in the case of the decay mode $H\to\gamma\gamma$.
In Ref.~\cite{Anastasiou:2007mz} the calculation was
extended to the decay mode $H\to WW\to l\nu l\nu$.
An independent NNLO calculation
of the Higgs production cross section has been presented in Refs.~\cite{Catani:2007vq,Grazzini:2008tf},
and implemented in the parton-level Monte Carlo program {\tt HNNLO}.
Such program allows the user to evaluate the Higgs production cross section with arbitrary kinematical cuts and
includes the decay $H\to\gamma\gamma$, $H\to WW\to l\nu l\nu$ and $H\to ZZ\to 4$ leptons.

Unfortunately, fixed order calculations may suffer from perturbative instabilities when different energy scales are involved.
An example is the transverse momentum $p_T$ spectrum of the Higgs boson.
In the small-$p_T$ region ($p_T\ll m_H$), the convergence
of the fixed-order expansion is spoiled by the presence of large logarithmic terms, $\alpha_S^n\ln^m(m_H^2/p_T^2)$.
To obtain reliable predictions, these logarithmically-enhanced terms have to be systematically
resummed to all perturbative orders \cite{Dokshitzer:hw}--\cite{Catani:2010pd}. It is then important to consistently
match the resummed and fixed-order calculations at intermediate values of $p_T$, in order to obtain
accurate QCD predictions for the entire range of transverse momenta.
In the case of Higgs boson production the resummation has been performed up to
next-to-next-to-leading logarithmic (NNLL) accuracy \cite{deFlorian:2000pr,Bozzi:2005wk} and matched
to the fixed order ${\cal O}(\as^4)$ result valid at large transverse momenta \cite{deFlorian:1999zd,Ravindran:2002dc,Glosser:2002gm}.
The calculation has been implemented in a numerical program, named {\tt HqT} \cite{grazzini_codes}
that has been used by the experimental collaborations at the Tevatron and the LHC for a few years.
In Ref.~\cite{Bozzi:2007pn} the calculation has been extended to include
the rapidity dependence of the Higgs boson.
% , but for technical reasons, such improvement was not implemented in {\tt HqT}.
In Ref.~\cite{deFlorian:2011xf}
we have improved the calculation of Ref.~\cite{Bozzi:2005wk} by
implementing the exact form of the NNLO coefficients of Ref.~\cite{Catani:2011kr}\footnote{The results of Ref.~\cite{Bozzi:2005wk,Bozzi:2007pn} were based on a reasonable approximation of these NNLO hard-collinear coefficients.} and
the value for the coefficient $A^{(3)}$ derived in Ref.~\cite{Becher:2010tm}.
The corresponding computation is implemented in a new version of {\tt HqT}.

In this paper we take one step forward with respect to the work
of Refs.~\cite{Bozzi:2005wk,Bozzi:2007pn,deFlorian:2011xf}.
We start from the doubly differential cross section, including transverse-momentum resummation and rapidity dependence \cite{Bozzi:2007pn} and we implement the hard collinear
coefficients of Ref.~\cite{Catani:2011kr} that, together with the exact form of the coefficient $A^{(3)}$ \cite{Becher:2010tm}, allow us to control the resummation at full NNLL accuracy. We then include the Higgs boson decay
and implement the ensuing result into an efficient Higgs event generator, that
is able to simulate the full kinematics of the Higgs boson
and of its decay products.
The resummed result is finally matched with the fixed order NNLO computation of Ref.~\cite{Catani:2007vq,Grazzini:2008tf} to obtain a prediction that is everywhere as good as the NNLO result, but
includes the resummation of the logarithmically enhanced contributions at small transverse momenta.
The exact form of the second order hard-collinear coefficients of Ref.~\cite{Catani:2011kr} permits a fully consistent matching
with the NNLO rapidity distribution upon integration over $p_T$.
The calculation is implemented in a new numerical program
called {\tt HRes}, that embodies the features of {\tt HNNLO} and {\tt HqT}.
We present a selection of numerical results that can be obtained with our program for Higgs boson production at the LHC ($\sqrt{s}=8$ TeV) up to NNLL+NNLO accuracy.
We consider the decay modes $H\to\gamma\gamma$, $H\to WW\to l\nu l\nu$ and $H\to ZZ\to 4l$ and we compare the resummed results with the corresponding
fixed order results, up to the NNLO accuracy,
obtained with the {\tt HNNLO} numerical code.

The paper is organized as follows. In Sect.~\ref{sec:hres} we
recall the main features of our resummation formalism
and we introduce
our NNLL+NNLO numerical program {\tt HRes}. In Sect.~\ref{sec:results} we present our numerical predictions at the LHC.
In Sect.~\ref{sec:summary} we summarize our results.

\section{Transverse momentum resummation and the {\tt HRes} program}
\label{sec:hres}

We start this Section by briefly recalling the resummation formalism of Refs.~\cite{Catani:2000vq,Bozzi:2005wk,Bozzi:2007pn}.
We consider the inclusive hard-scattering process
\begin{equation}
\label{process}
h_1(p_1) + h_2(p_2) \to H(y,p_T,m_H) + X \;\;,
\end{equation}
where the collision of the two hadrons $h_1$ and $h_2$ with
momenta $p_1$ and $p_2$ produces the Higgs boson $H$ with transverse momentum $p_T$ and rapidity $y$ (defined in the centre-of-mass frame) accompanied by an arbitrary and undetected final state $X$.
The centre-of-mass energy of the colliding hadrons is denoted by $\sqrt s$.
%$(s= (p_1+p_2)^2 \simeq 2p_1p_2)$.
%The rapidity, $y$, of the Higgs boson is defined in the centre-of-mass frame
%of the colliding hadrons, and the forward direction ($y > 0$) is identified by 
%the direction of the momentum $p_1$.

According to the QCD factorization theorem,
the doubly differential cross 
section for this process reads
\begin{eqnarray}
\f{d\sigma}{dy \, dp_T^2}(y,p_T,m_H,s) &=& \sum_{a_1,a_2}
\int_0^1 dx_1 \,\int_0^1 dx_2 \; f_{a_1/h_1}(x_1,\mu_F^2)
\,f_{a_2/h_2}(x_2,\mu_F^2) \nn \\
\label{dcross}
&\times&
\f{d{\hat \sigma}_{a_1a_2}}{d{\hat y} \,dp_T^2}({\hat y},p_T, m_H,{\hat s};
\as(\mu_R^2),\mu_R^2,\mu_F^2) 
\;,
\end{eqnarray}
where $f_{a/h}(x,\mu_F^2)$ ($a=q_f,{\bar q_f},g$) are the parton densities of 
the colliding hadrons at the factorization scale $\mu_F$,
$d{\hat \sigma}_{ab}$ are the
partonic cross sections, and $\mu_R$ is the renormalization scale.
%Throughout the paper we use parton
%densities as defined in the \ms\
%factorization scheme, and $\as(q^2)$ is the QCD running coupling in the \ms\
%renormalization scheme.
The rapidity, $\hat y$, and the centre-of-mass energy,
${\hat s}$, of the partonic cross section (subprocess) are related to
the corresponding hadronic variables $y$ and $s$ as:
\begin{equation}
\label{kin}
{\hat y} = y - \frac{1}{2} \ln\frac{x_1}{x_2} \;, \quad 
\quad {\hat s}=x_1x_2s \;\;.
\end{equation}
%with the kinematical boundary $|{\hat y}| < \ln \sqrt{{\hat s}/M^2}$
%$(\, |y| < \ln \sqrt{s/M^2} \,)$ and ${\hat s} > M^2$ $(s > M^2)$.

The partonic cross section $d{\hat \sigma}_{ab}$ is computable in QCD 
perturbation theory but its series expansion in $\as$ contains  
the logarithmically-enhanced terms, $(\as^n/p_T^2)\, \ln^m (m_H^2/p_T^2)$,
that we want to resum.

To this purpose, the partonic cross section is rewritten as the sum of two terms,
\begin{equation}
\label{resfin}
\f{d{\hat \sigma}_{a_1a_2}}{d{\hat y} \,dp_T^2} =
\f{d{\hat \sigma}_{a_1a_2}^{(\rm res.)}}{d{\hat y} \,dp_T^2}
+\f{d{\hat \sigma}_{a_1a_2}^{(\rm fin.)}}{d{\hat y} \,dp_T^2} \;\;.
\end{equation}
The logarithmically-enhanced contributions are embodied in the 
`resummed' component $d{\hat \sigma}_{a_1a_2}^{(\rm res.)}$.
The `finite' component $d{\hat \sigma}_{a_1a_2}^{(\rm fin.)}$
is free of such contributions, and it
can be computed by 
a truncation 
of the perturbative series at a given fixed order.
In particular we compute $d{\hat \sigma}_{a_1a_2}^{(\rm fin.)}$
starting from $\left[d{\hat \sigma}_{a_1a_2}\right]_{\rm f.o.}$,
the usual perturbative series truncated at a given fixed order in $\as$,
and we subtract the perturbative truncation of the resummed component at
the same order:
\begin{equation}
\left[\f{d{\hat \sigma}_{a_1a_2}^{(\rm fin.)}}{d{\hat y} \,dp_T^2}\right]_{\rm f.o.}=\Bigg[\f{d{\hat \sigma}_{a_1a_2}}{d{\hat y} \,dp_T^2}\Bigg]_{\rm f.o.}-
\left[\f{d{\hat \sigma}_{a_1a_2}^{(\rm res.)}}{d{\hat y} \,dp_T^2}\right]_{\rm f.o.}\, .
\end{equation}

The resummation procedure of the logarithmic terms has to be carried out 
\cite{Parisi:1979se}-\cite{Kodaira:1981nh}
in the impact-parameter space, to correctly take into account the 
kinematics constraint of transverse-momentum conservation.
The resummed component of the partonic cross section 
is then obtained by performing the inverse Fourier (Bessel) transformation 
with respect to the impact parameter $b$.
We write\footnote{In the following equations, the functional dependence on 
the scales $\mu_R$ and $\mu_F$ is understood.}
\beq
\label{resum}
%\left[\f{d{\hat \sigma}_{ab}}{d p_T^2}\right]_{\rm res.}=
\!\!\! \f{d{\hat \sigma}_{a_1a_2}^{(\rm res.)}}{d{\hat y} \,dp_T^2}({\hat y}, 
p_T,m_H,{\hat s};
\as)
%\as(\mu_R^2),\mu_R^2,\mu_F^2) 
%%%%%
%&=& \f{m^2_H}{\hat s} \;\int \f{d^2{\bf b}}{4\pi} \;e^{i {\bf b} \cdot 
%{\bf \qt}}
%\;{\cal W}_{a_1a_2}({\hat y},b,m_H,{\hat s};\as)
%%\as(\mu_R^2),\mu_R^2,\mu_F^2)  
%\\
%\label{resum1}
= \f{m^2_H}{\hat s} \;
\int_0^\infty db \; \f{b}{2} \;J_0(b p_T) 
\;{\cal W}_{a_1a_2}({\hat y},b,m_H,{\hat s};\as)
%\as(\mu_R^2),\mu_R^2,\mu_F^2) 
\;,
\eeq
where $J_0(x)$ is the 0th-order Bessel function, and the
factor ${\cal W}$
embodies the all-order dependence on 
the large logarithms $\ln (m_H^2b^2)$ at large $b$, which correspond to
$\ln (m^2_H/p_T^2)$ terms in $p_T$ space.

In the case of the $p_T$ cross section integrated over the rapidity,
the resummation of the large logarithms
is better expressed \cite{Catani:2000vq,Bozzi:2005wk}
by defining the $N$-moments ${\cal W}_N$ of ${\cal W}$
with respect to $z=m^2_H/{\hat s}$ at fixed $m_H$. In the present case, in which we want to keep the dependence on the rapidity into account, we
consider `double' $(N_1,N_2)$-moments
with respect to the two variables
$z_1=e^{+{\hat y}} m_H/{\sqrt{\hat s}}$ and 
$z_2=e^{-{\hat y}} m_H/{\sqrt{\hat s}}$ at fixed $m_H$
(note that $0< z_i <1$).
We thus introduce ${\cal W}^{(N_1,N_2)}$ as follows \cite{Bozzi:2007pn}:
\begin{equation}
\label{wnnudef}
{\cal W}_{a_1a_2}^{(N_1,N_2)}(b,m_H;\as) =
\int_0^1 dz_1 \,z_1^{N_1-1} \; \int_0^1 dz_2 \,z_2^{N_2-1} \;\, 
{\cal W}_{a_1a_2}({\hat y},b,m_H,
{\hat s};\as)
\;.
\end{equation}

The convolution structure of the QCD factorization formula
(\ref{dcross}) is easily diagonalized by considering $(N_1,N_2)$-moments:
\begin{equation}
\label{n12fact}
d\sigma^{(N_1,N_2)} = \sum_{a_1,a_2} \;f_{a_1/h_1, N_1+1} \;f_{a_2/h_2,N_2+1}
\;d{\hat \sigma}_{a_1a_2}^{(N_1,N_2)} \;,
\end{equation}
where $f_{a/h, N}= \int_0^1 dx \,x^{N-1} f_{a/h}(x)$ are the customary
$N$-moments of the parton distributions.

The use of Mellin moments also simplifies the resummation structure of the
logarithmic terms in  
$d{\hat \sigma}_{a_1a_2}^{({\rm res.}) \,(N_1,N_2)}$.
The perturbative factor ${\cal W}_{a_1a_2}^{(N_1,N_2)}$
can indeed be organized in exponential form as follows:
\begin{equation}
\label{wtilde}
{\cal W}^{(N_1,N_2)}(b,m_H;\as) =
{\cal H}^{(N_1,N_2)}(m_H,\as;m_H^2/Q^2) 
\; \exp\{{\cal G}^{(N_1,N_2)}(\as,{\widetilde L};m_H^2/Q^2)\}
\;\;,
\end{equation}
where 
\begin{equation}
\label{logdef}
{\widetilde L}= \ln\left(\f{Q^2\,b^2}{b_0^2} + 1\right) \;\;,
\end{equation}
$b_0=2e^{-\gamma_E}$ 
($\gamma_E=0.5772\dots$ is the Euler number) and, to simplify the notation,
the dependence on the flavour indeces has been understood.
The scale $Q\sim m_H$ in Eq.~(\ref{logdef}), named resummation scale, parametrizes the arbitrariness in the resummation procedure. Its role is analogous
to the role played by the renormalization (factorization) scale in the context
of the renormalization (factorization) procedure. Although the resummed
cross section does not depend on $Q$ when evaluated at all perturbative orders,
its explicit dependence on $Q$ appears after truncation of the resummed expression at a given logarithmic accuracy.

The function ${\cal H}^{(N_1,N_2)}$
does not depend on the impact parameter $b$ and, therefore, its evaluation
does not require resummation of large logarithmic terms. It can be expanded
in powers of $\as$ as
\begin{equation}
\label{hexpan}
{\cal H}^{(N_1,N_2)}=
\sigma_{0}(\as,m_H)
\Bigl[ 1+ \f{\as}{\pi} \,{\cal H}^{(N_1,N_2) \,(1)} +
\left(\f{\as}{\pi}\right)^2 
\,{\cal H}^{(N_1,N_2) \,(2)}
+ \dots \Bigr]\;\;, 
\end{equation}
where $\sigma_{0}(\as,m_H)$ is the lowest-order partonic cross section for
Higgs boson production.
The form factor $\exp\{{\cal G}\}$
includes the complete dependence on $b$ and,
in particular, it contains 
all the terms that order-by-order in $\as$ are logarithmically
divergent when $b \to \infty$. The functional dependence on $b$ is 
expressed through 
the large logarithmic terms $\as^n {\widetilde L}^m$
with $1\leq m \leq 2n$. 
More importantly,
all the logarithmic contributions to ${\cal G}$ with $n+2 \leq m \leq 2n$
are vanishing.
Thus, the exponent ${\cal G}$ can systematically
be expanded in powers of $\as$, at fixed value of $\lambda=\as {\widetilde L}$,
as follows:
\begin{equation}
\label{gexpan}
{\cal G}^{(N_1,N_2)}(\as,{\widetilde L};m_H^2/Q^2) =
{\widetilde L} \,g^{(1)}(\as {\widetilde L})+
g^{(2) \,(N_1,N_2)}(\as {\widetilde L};m_H^2/Q^2) +
\f{\as}{\pi} \;g^{(3) \,(N_1,N_2)}(\as {\widetilde L};m_H^2/Q^2) + \dots
\;\;. 
\end{equation}
The term ${\widetilde L} g^{(1)}$ collects the leading logarithmic (LL) 
contributions $\as^n {\widetilde L}^{n+1}$;
the function $g^{(2)}$ includes
the next-to-leading logarithmic (NLL) contributions 
$\as^n {\widetilde L}^{n}$; 
$g^{(3)}$ resums the next-to-next-to-leading logarithmic (NNLL) terms
$\as^n {\widetilde L}^{n-1}$, and so forth.  

Note that we use the logarithmic variable ${\widetilde L}$ 
(see Eq.~(\ref{logdef}))
to organize the resummation of the large logarithms $\ln (Q^2b^2)$.
In the region in which $Qb \gg 1$ we have
${\widetilde L} \sim \ln (Q^2b^2)$ and the use of the variable 
${\widetilde L}$ is fully legitimate to arbitrary logarithmic accuracy.
When $Qb \ll 1$, we have $\tL \to 0$ 
%(whereas $\ln (Q^2b^2) \to \infty$ !)
and $\exp \{{\cal G}(\as, \tL)\} \to 1$.
Therefore, the use of ${\widetilde L}$ reduces the
effect produced by the resummed contributions
in the small-$b$ region (i.e., at large and intermediate values of $p_T$), 
where the large-$b$ resummation approach is not justified.
In particular, 
setting $b=0$ (which corresponds to integrate over the entire $p_T$ range)
we have $\exp \{{\cal G}(\as, \tL)\} = 1$: this property
can be interpreted \cite{Bozzi:2005wk}
as a unitarity constraint on the total cross section;
transverse-momentum resummation smears the shape of the
$p_T$ distribution of the Higgs boson
without affecting its total production rate.

The resummation formulae (\ref{wtilde}), (\ref{hexpan})
and (\ref{gexpan}) can be worked out
at any logarithmic accuracy
since the functions ${\cal H}$ and ${\cal G}$ can 
be expressed (see Refs.~\cite{Bozzi:2005wk,Catani:2010pd}) 
in terms of few perturbatively-computable coefficients 
denoted by $A^{(n)}, B^{(n)},
H^{(n)}, C^{(n)}_N, G^{(n)}_N, \gamma^{(n)}_N$.
In the case of the $p_T$ cross section integrated over the rapidity, 
Eq.~(\ref{wtilde}) is still valid, provided the double 
$(N_1,N_2)$-moments are replaced by the corresponding single $N$-moments
${\cal W}_N, {\cal H}_N, {\cal G}_N$ (see Sect.~2.2 in 
Ref.~\cite{Bozzi:2005wk} and Sect.~2 in Ref.~\cite{Bozzi:2007pn}).

The formalism briefly recalled in this section defines 
a systematic expansion \cite{Bozzi:2005wk} of Eq.~(\ref{resfin}):
it can be used to obtain predictions that, formally, have uniform perturbative
accuracy from the small-$p_T$ region to the large-$p_T$ region. The various
orders of this expansion are denoted
as NLL+NLO, NNLL+NNLO, etc., where the 
first label (NLL, NNLL, $\dots$) refers to the logarithmic accuracy at 
small $p_T$ and the second label (NLO, NNLO, $\dots$) refers to the customary 
perturbative order for the inclusive cross section\footnote{We note that
this notation differs from the one used
in Refs.~\cite{Bozzi:2005wk,Bozzi:2007pn,deFlorian:2011xf}, where the various terms of the expansion were denoted by NLL+LO, NNLL+NLO and so forth.
Since in this paper we do not limit ourselves to study the Higgs $p_T$ spectrum, we prefer to label the fixed order contributions entering the matching procedure according to the order they contribute to the inclusive cross section.}. 
To be precise,
the NLL+NLO term of Eq.~(\ref{resfin}) is obtained by including  
the functions $g^{(1)}$, $g^{(2)}$ \cite{Catani:vd}
and the coefficient ${\cal H}^{(1)}$ \cite{Kauffman:1991cx,Yuan:1991we} (see Eqs.~(\ref{gexpan}) and 
(\ref{hexpan}))
in the resummed component, and by expanding the finite (i.e. large-$p_T$)
component up to ${\cal O}(\as^3)$. At NNLL+NNLO accuracy, the resummed component
includes also the function $g_N^{(3)}$ \cite{deFlorian:2000pr} and the coefficient 
${\cal H}^{(2)}$ \cite{Catani:2011kr} (see Eqs.~(\ref{gexpan}) and 
(\ref{hexpan})), while the finite component is expanded up to ${\cal O}(\as^4)$.
We point out
that the NNLL+NNLO (NLL+NLO) result includes the {\em full} NNLO (NLO)
perturbative
contribution, supplemented with the resummation of the
logarithmically enhanced terms in the small-$p_T$ region at (N)NLL.

In order implement our calculation in a tool that can be used to perform realistic simulations,
it is important to consider the Higgs boson decays.
Since we are dealing with a scalar particle, the inclusion of the Higgs decay does not lead to substantial complications. However, the efficient generation of ``Higgs events'' according to the doubly-differential distribution
of Eq.~(\ref{resfin}) and the inclusion of the decay are technically non trivial and require substantial improvements
in the speed of the numerical program that evaluates the resummed cross section. 
The finite part in Eq.~(\ref{resfin}) is instead evaluated through an appropriate modification of the {\tt HNNLO} code, which
being based on the subtraction formalism of Ref.~\cite{Catani:2007vq}, is particularly suitable to this purpose.

We recall \cite{Bozzi:2005wk} that, due to
our actual definition of the logarithmic parameter ${\widetilde L}$ in
Eq.~(\ref{wtilde}) and to our matching procedure with the perturbative 
expansion at large $p_T$, the integral over $p_T$ of the $p_T$ cross section
exactly reproduces the customary fixed-order calculation of the total cross
section. This integral, however, implies an extrapolation of the resummed
result at large transverse momenta, where the resummation cannot improve the accuracy of the fixed order expansion.
Moreover, the extrapolation of the resummed cross section at large transverse momenta may lead to
unjustified large uncertainties and ensuing lack of predictivity
(see Sect.~3 in Ref.~\cite{Bozzi:2005wk}).
This is not a problem if the calculation is limited to the transverse momentum spectrum. In this case, in fact, we
can simply use the fixed order result when the uncertainty of the resummed calculation becomes too large.
In the present case, since our goal is to generate the full kinematics of the Higgs boson and its decays
without a selection on the Higgs transverse momentum, this issue becomes particularly relevant.
In the numerical implementation of Eq.~(\ref{resfin}) we thus introduce a smooth
switching procedure at large $p_T$, by replacing
the resummed cross section in Eq.~(\ref{resfin}) as follows:
\begin{equation}
\label{switch}
\f{d{\hat \sigma}_{a_1a_2}}{d{\hat y} \,dp_T^2} \to w(p_T)
\left(\f{d{\hat \sigma}_{a_1a_2}^{(\rm res.)}}{d{\hat y} \,dp_T^2}
+\f{d{\hat \sigma}_{a_1a_2}^{(\rm fin.)}}{d{\hat y} \,dp_T^2}\right)
+(1-w(p_T))\Bigg[\f{d{\hat \sigma}_{a_1a_2}}{d{\hat y} \,dp_T^2}\Bigg]_{\rm f.o.}\, .
\end{equation}
where the function $w(p_T)$ is defined as\footnote{We note that a simpler switching option is available in the new version of {\tt HqT} \cite{grazzini_codes}.}
\begin{equation}
w(p_T) =
\bigg \{
\begin{array}{cl}
1 & p_T \leq p_T^{\rm sw.}-\Delta p_T \\
f(p_T) & p_T^{\rm sw.}-\Delta p_T< p_T < p_T ^{\rm sw.}+\Delta p_T \\
0 & p_T \geq p_T^{\rm sw.}+\Delta p_T \\
\end{array}
\end{equation}
and the function $f(p_T)$ is chosen in such a way that $w(p_T)$ and $w^\prime(p_T)$ are continuous in all the range of transverse momenta.
In particular, we choose
\begin{equation}
\label{fswitch}
f(p_T)=\f{1}{2}\left(\cos\left(\pi\f{p_T-(p_T^{\rm sw.}-\Delta p_T)}{2\Delta p_T}\right)+1\right)\, .
\end{equation}
We have checked that the parameters $p_T^{\rm sw.}$ and $\Delta p_T$ can be consistently chosen
so as not to spoil our unitarity constraint,
and that the integral of our NLL+NLO and NNLL+NNLO
resummed result still reproduces well the NLO and NNLO inclusive cross
sections (see Sec.~3).

We have implemented our calculation in a numerical program called {\tt HRes},
by considering three decay modes: $H\to\gamma\gamma$, $H\to WW\to l\nu l\nu$
and $H\to ZZ\to 4$ leptons. In the latter case the user can choose between $H\to ZZ\to \mu^+\mu^- e^+e^-$ and $H\to ZZ\to e^+e^-e^+e^-$, which includes the appropriate interference contribution.
The program can be downloaded from \cite{grazzini_codes}, together with
some accompanying notes.

\section{Results}
\label{sec:results}

\subsection{Preliminaries}
\label{sec:decay}

We consider Higgs boson production at the LHC (e.g. $pp$ collisions at $\sqrt{s}=8$ TeV).  In order to avoid a multiple presentation of similar results we
use MSTW2008 parton distributions \cite{Martin:2009iq},
with densities and $\as$ evaluated at each corresponding order,
i.e. we use $(n+1)$-loop $\as$ at N$^n$LL+N$^n$LO and N$^n$LO (with $n=1,2$), and 1-loop $\as$ for LO.
Unless stated otherwise, renormalization, factorization and resummation scales
are set to their default values, $\mu_R=\mu_F=2Q=m_H$.
We remind the reader that the calculation is performed in the $M_t\to \infty$ limit.

As for the electroweak couplings, we use the scheme where
the input parameters are $G_F$, $m_Z$, $m_W$ and $\alpha(m_Z)$.
In particular we take $G_F=1.16639\times 10^{-5}$ GeV$^{-2}$, $m_Z=91.188$ GeV, $m_W=80.419$ GeV and $\alpha(m_Z)=1/128.89$.
The decay matrix elements are implemented at Born level, i.e.,
radiative corrections are completely neglected.
The Higgs boson is treated in the narrow-width approximation,
but in the $W$ and $Z$ decays we take into account finite width effects,
by using $\Gamma_W=2.06$ GeV and $\Gamma_Z=2.49$ GeV.
As explained in Sect.~\ref{sec:hres}, in order to obtain meaningful predictions in the entire range of transverse momenta,
we apply a smooth switching procedure (see Eq.~(\ref{switch})).
In our numerical implementation the parameters in Eq.~(\ref{fswitch}) are phenomenologically chosen to be $\Delta p_T=30$ GeV and $p_T^{\rm sw}=a\,m_H+b(m_H/Q+c)\sqrt{s}$.
At NLL+NLO accuracy we set $a=1/2$, $b=1.2\times 10^{-3}$ and $c=2.5$, whereas 
at NNLL+NNLO we set $a=0.6$, $b=2.8\times 10^{-3}$ and $c=0$.
We postpone some comments on the dependence of our results on these parameters to Sect.~\ref{sec:discuss}.

\subsection{$H\to \gamma\gamma$}

We first consider the production of a SM Higgs boson with mass $m_H=125$ GeV.
The width is computed with the program HDECAY \cite{Djouadi:1997yw} to be $\Gamma_H=4.15$ MeV.
With this choice of $m_H$, the effects of finite width can safely be neglected.

\begin{figure}[!ht]
\centering
\includegraphics[width=0.7\textwidth]{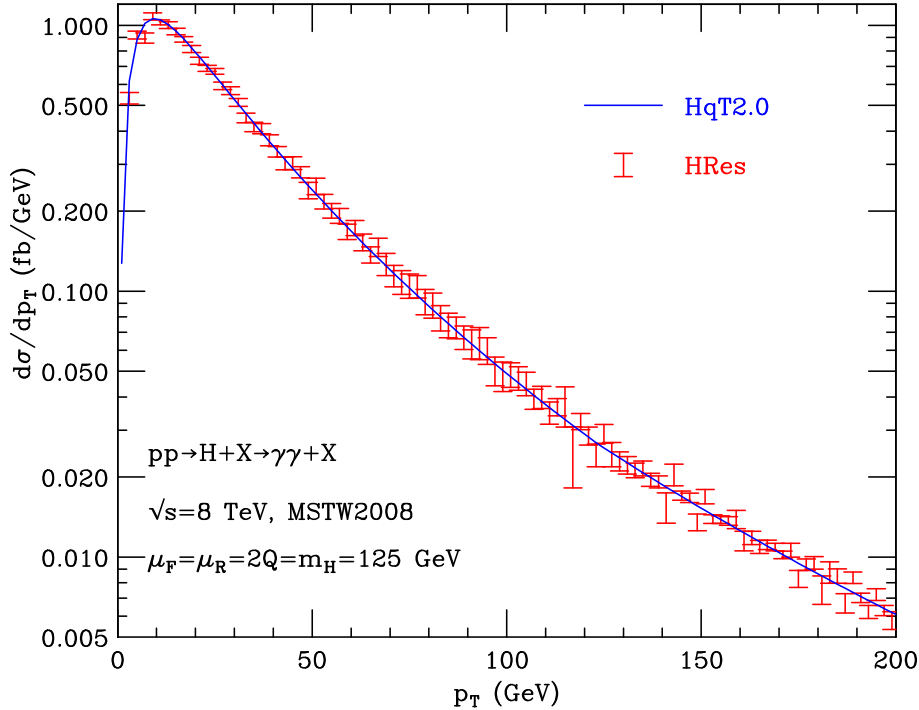}
\caption{{\em Transverse momentum spectrum for the $H\to\gamma\gamma$ signal
at the LHC for $m_H=125$ GeV, obtained at NNLL+NNLO with {\tt HRes} compared to the corresponding result from {\tt HqT}. The result from {\tt HqT} is multiplied by the branching ratio $BR(H\to\gamma\gamma)=2.245\times 10^{-3}$ \cite{Djouadi:1997yw}.}}
\label{fig:pTH}
\end{figure}
When no cuts are applied, the $p_T$ spectrum of the Higgs boson obtained with {\tt HRes} must be in agreement with the one obtained with the {\tt HqT} numerical program. In Fig.~\ref{fig:pTH} we compare the two spectra to check that this is indeed the case, within the statistical uncertainties.
The corresponding inclusive cross sections are reported in Table \ref{tab:gamgacentralscale}, where we show the new resummed results obtained through the {\tt HRes} code, and we compare them with the fixed order predictions obtained with the {\tt HNNLO} code. We see that the NLL+NLO (NNLL+NNLO) inclusive cross section agrees with the NLO (NNLO) result to better than $1\%$.

As an example, we apply the following cuts on the photons. For each event, we classify the photon transverse momenta according to their minimum and maximum value, $p_{T\rm{min}}$ and $p_{T\rm{max}}$ . The photons are required to be in the central rapidity region, $|\eta|<2.5$, with $p_{T\rm{min}}>25$ GeV and $p_{T\rm{max}}>40$ GeV.
Note that an isolation cut on the photons is generally required.
For example, a standard isolation is to require the total transverse energy
in a cone of a given radius $R$ around each photon to be smaller than
a fraction of the photon $p_T$. 
Such cuts cannot be taken into account in our resummed
calculation, since we are inclusive over the QCD radiation recoiling against the Higgs boson. Their effect can be estimated with the {\tt HNNLO} code and turns out to be rather small.

\begin{table}[!ht]
\begin{center}
\begin{tabular}{|c||c|c|c|c|}
\hline
Cross section & NLO & NLL+NLO & NNLO & NNLL+NNLO\\ \hline \hline
Total [fb]& 30.65 $\pm$ 0.01 & 30.79 $\pm$ 0.03 & 38.47 $\pm$ 0.15 & 38.41 $\pm$ 0.06\\ \hline
With cuts [fb]& 21.53 $\pm$ 0.02 & 21.55 $\pm$ 0.01 & 27.08 $\pm$ 0.08 & 26.96 $\pm$ 0.04\\ \hline
Efficiency [\%] & 70.2 & 70.0 & 70.4 & 70.2\\
\hline
\end{tabular}
\end{center}
\caption{{\em Fixed order and resummed cross sections for $pp\to H+X\to \gamma\gamma+X$ at the LHC, before and after geometrical acceptance cuts.}}
\label{tab:gamgacentralscale}
\end{table}

We recall that the resummation does not affect the total cross section for the Higgs boson production, but when geometrical cuts are applied, their effect can act in a different way on fixed order and resummed calculations. In Table \ref{tab:gamgacentralscale} we compare the accepted cross sections, obtained by the fixed order and resummed calculations,
and the corresponding efficiencies. 
The numerical errors estimate the statistical uncertainty of the Monte Carlo integration.
Comparing resummed and fixed order predictions, we see that there are no substantial differences on the accepted cross section, due to the fact that the integration is performed over a wide kinematical range. In Table \ref{tab:cut_gamma} we report the accepted cross section for different choices of the scales. After selection cuts, the scale uncertainty is about $\pm 15\%$ ($\pm 18 \%$) at NLL+NLO (NLO) and $\pm 9\%$ ($\pm 10\%$) at NNLL+NNLO (NNLO).

\begin{table}[!ht]
\vspace{0.3cm}
\centering
\begin{tabular}{|c||c|c|c|c|}
\hline
Cross section [fb] & NLO & NLL+NLO & NNLO & NNLL+NNLO\\ \hline \hline
$(2Q=\mu_F=\mu_R)=m_H/2$ & 25.92 $\pm$ 0.02 & 25.57 $\pm$ 0.03 & 29.52 $\pm$ 0.13 & 29.59 $\pm$ 0.11 \\ \hline
$(2Q=\mu_F=\mu_R)=m_H$ & 21.53 $\pm$ 0.02 & 21.55 $\pm$ 0.01 & 27.08 $\pm$ 0.08 & 26.96 $\pm$ 0.04\\ \hline
$(2Q=\mu_F=\mu_R)=2m_H$ & 18.17 $\pm$ 0.01 & 18.80 $\pm$ 0.02 & 24.43 $\pm$ 0.06 & 24.69 $\pm$ 0.06 \\ \hline
\end{tabular}
\caption{{\em NLO, NLL+NLO, NNLO and NNLL+NNLO accepted cross sections for $pp\to H+X\to \gamma\gamma+X$ at the LHC, for different choices of the scales.}}
\label{tab:cut_gamma}
\end{table}

\begin{figure}[!ht]
\centering
\includegraphics[width=0.65\textwidth]{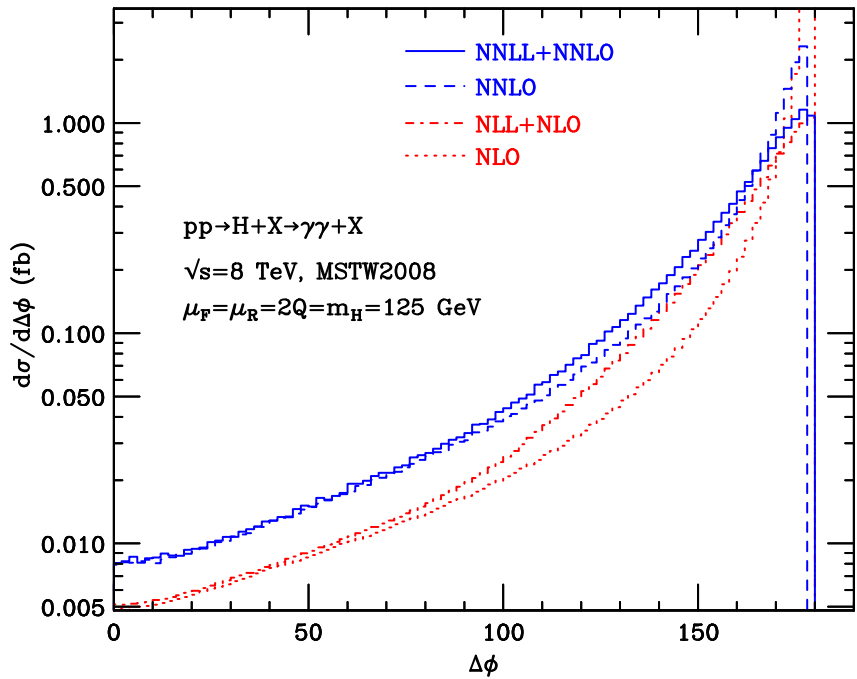}
\caption{{\em $\Delta\phi$ distribution from the $H\to\gamma\gamma$ signal at the LHC, obtained by the fixed order and resummed calculation.}}
\label{fig:dphi_photons}
\end{figure}

In Fig.~\ref{fig:dphi_photons} we study the distribution in the
azimuthal separation of the photons in the transverse plane, $\Delta\phi$.
At LO the photons are back-to-back, and thus $\Delta\phi$ is 180\textdegree.
Beyond LO, events with $\Delta\phi$ different from 180\textdegree\ are allowed, but NLO (dots) and NNLO (dashes) results show an unphysical behaviour as $\Delta\phi\to 180^o$.
The resummed NLL+NLO and NNLL+NNLO results lead instead to a smooth behaviour in this region. On the other hand, $\Delta\phi\to 0$ corresponds to a kinematical configuration where the diphoton system is produced with large transverse momentum, so the result is fully dominated by the corresponding fixed order calculation. 

An interesting observable, which has been used by ATLAS to split the $H\to \gamma\gamma$
analysis in categories \cite{:2012sk},
is the thrust transverse momentum $p_{Tt}$\footnote{In the context of Drell-Yan lepton pair production, this variable is also called $a_T$ \cite{Vesterinen:2008hx,Banfi:2009dy}.} \cite{Vesterinen:2008hx}. Defining the thrust axis $\hat t$ and the transverse momentum of the diphoton system $\vec p_{\rm T}^{\;\gamma\gamma}$ as follows
\begin{eqnarray}
\hat t=\frac{\vec p_{\rm T}^{\;\gamma_1}-\vec p_{\rm T}^{\;\gamma_2}}{|\vec p_{\rm T}^{\;\gamma_1}-\vec p_{\rm T}^{\;\gamma_2}|};&&\vec p_{\rm T}^{\;\gamma\gamma}=\vec p_{\rm T}^{\;\gamma_1}+\vec p_{\rm T}^{\;\gamma_2},
\end{eqnarray}
the $p_{Tt}$ is then calculated according to:
\begin{equation}
p_{Tt}=|\vec p_{\rm T}^{\;\gamma\gamma}\times \hat t|.
\label{eq:ptt}
\end{equation}

\begin{figure}[!ht]
\centering
\includegraphics[width=0.7\textwidth]{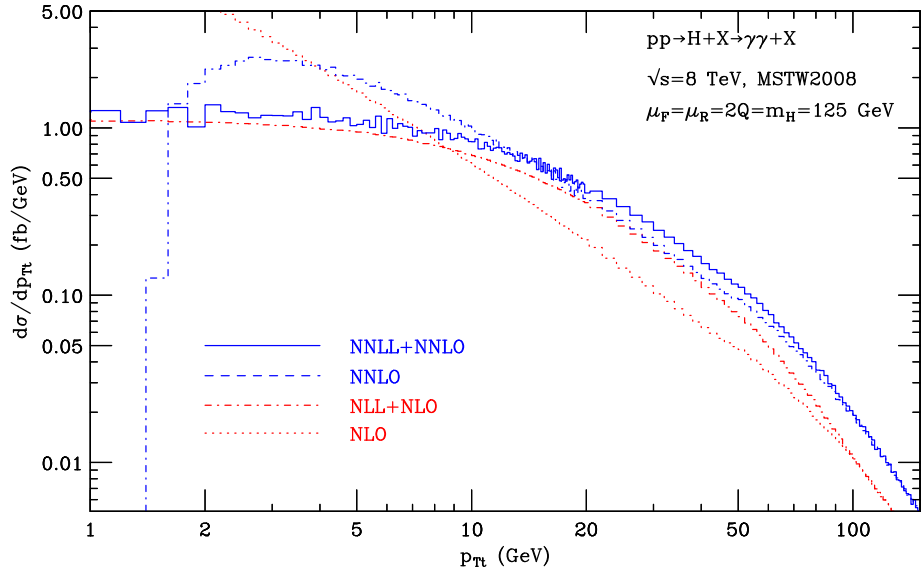}
\caption{{\em $p_{Tt}$ distribution for the $H\to\gamma\gamma$ signal at the LHC, obtained at NLL+NLO and NNLL+NNLO compared to the corresponding NLO and NNLO results.}}
\label{fig:pTthrust}
\end{figure}

In Fig. \ref{fig:pTthrust} we report the $p_{Tt}$ distribution, obtained
at NLO (dots), NNLO (dashes), NLL+NLO (dot dashes) and NNLL+NNLO (solid).
We see that in the high $p_{Tt}$ region the NLL+NLO prediction agrees with the NLO one, and the NNLL+NNLO prediction agrees with NNLO. In the low $p_{Tt}$ region the NLO result diverges to $+\infty$, whereas the NNLO diverges
to $-\infty$. Such behaviour is analogous to the behaviour of the $p_T$ distribution of the Higgs boson when computed at fixed order in QCD perturbation theory.
The NLL+NLO and NNLL+NNLO results obtained with {\tt HRes} are instead finite as $p_{Tt}\to 0$, approaching a constant value.

\begin{figure}[!ht]
\begin{center}
\subfigure[]{\label{fig1a}
\includegraphics[width=0.485\textwidth]{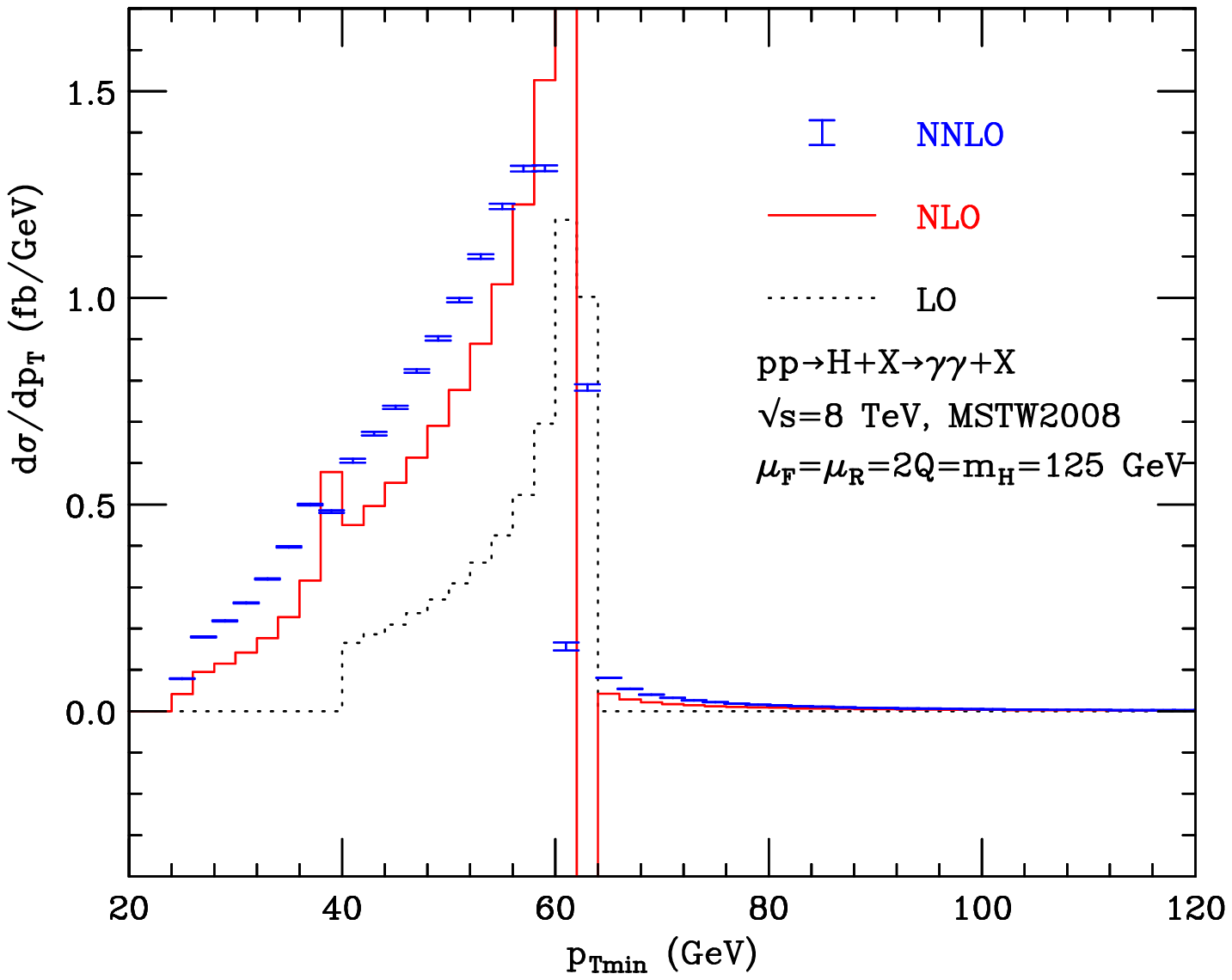}}
\subfigure[]{\label{fig1b}
\includegraphics[width=0.485\textwidth]{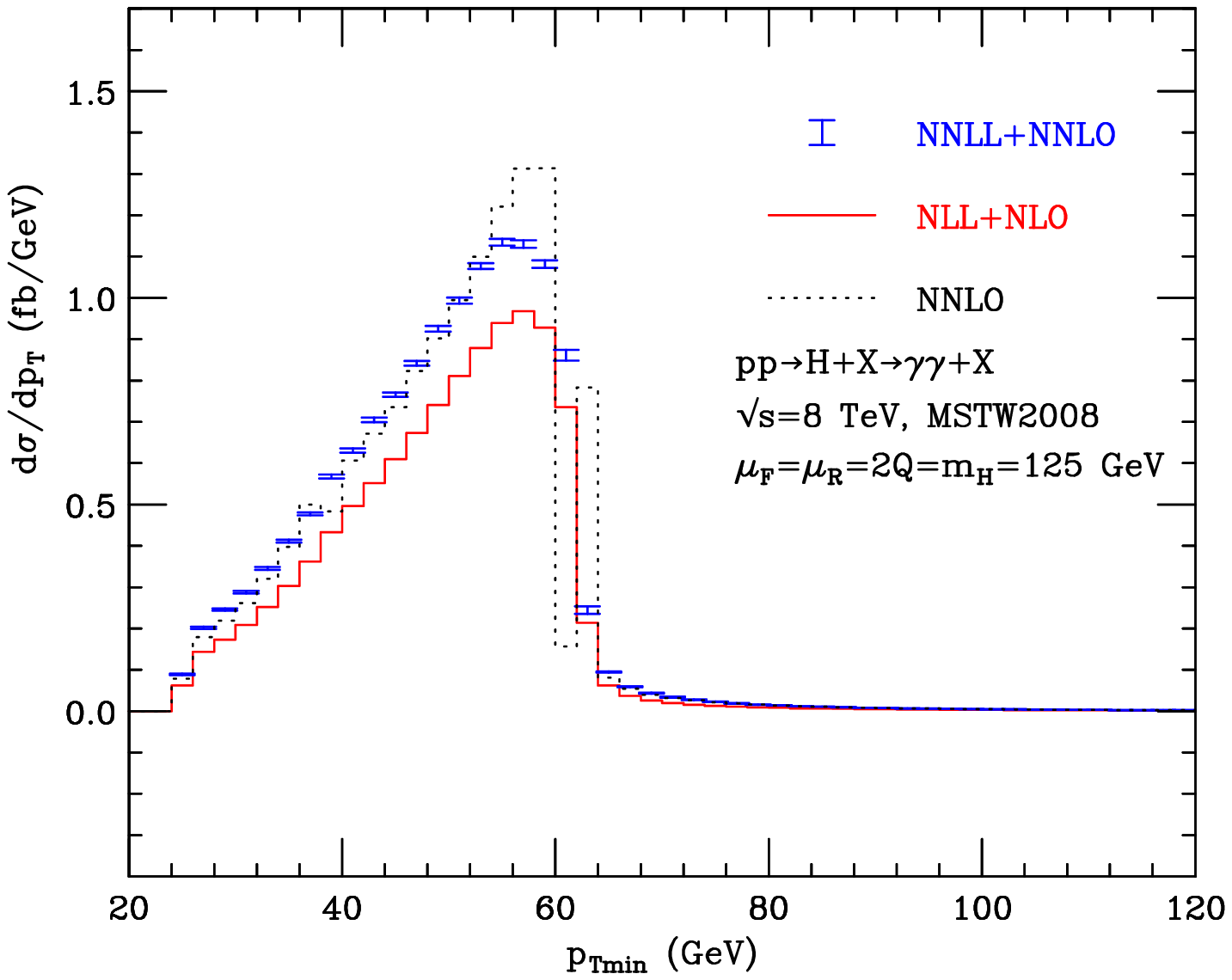}}
\subfigure[]{\label{fig1c}
\includegraphics[width=0.485\textwidth]{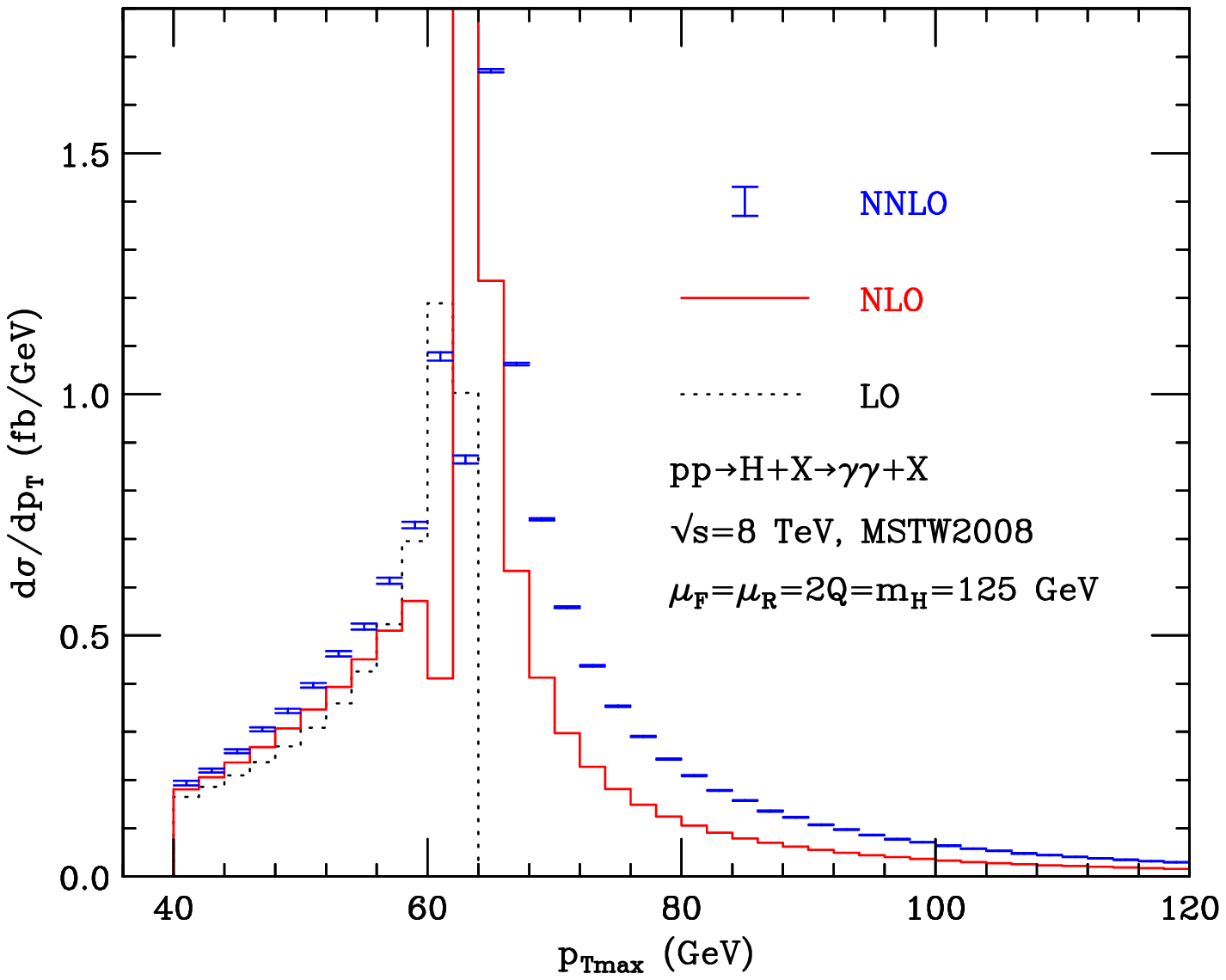}}
\subfigure[]{\label{fig1d}
\includegraphics[width=0.485\textwidth]{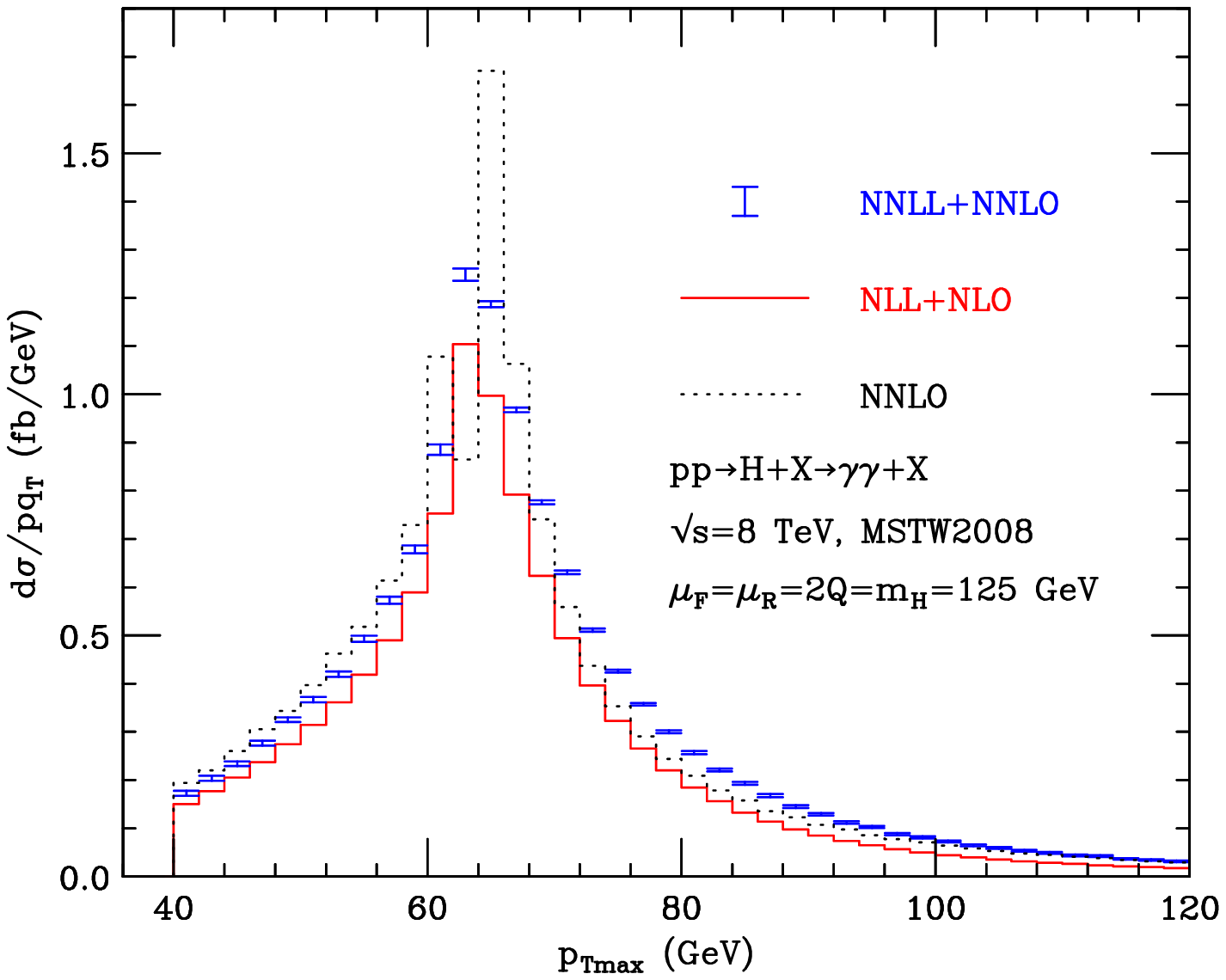}}
\vspace*{-.5cm}
\end{center}
\caption{\label{fig:pTphoton}{\em Distributions in $\ptmin$ (a,b) and $\ptmax$ (c,d) for the
$H\to \gamma\gamma$ signal at the LHC, obtained by fixed order (a,c) and resummed (b,d) calculations. In the right panels the fixed order NNLO result is also shown for comparison.}}

\end{figure}

In Fig. \ref{fig1a},\ref{fig1c} we plot the photon $p_T$ distributions $\ptmin$ and $\ptmax$. These distribution are enhanced when going from LO to NLO to NNLO according to the increase of the total cross section. We note that, as pointed out in Ref.~\cite{Catani:2007vq}, the shape of these distributions sizeable differs when going from LO to NLO and to
NNLO. In particular, at the LO the two photons are emitted with the same $p_T$ because the Higgs boson is produced with zero transverse momentum, hence the LO $\ptmin$ and $\ptmax$ are exactly identical. Furthermore the LO distribution has a kinematical boundary at $p_T = m_H /2$ (Jacobian peak), which is due to the use of the narrow width approximation. Such condition is released once extra radiation is accounted for. Thus higher order predictions suffer of perturbative instabilities, i.e. each higher-order perturbative contribution produces (integrable) logarithmic singularities in the vicinity of that boundary, as explained in Ref.~\cite{Catani:1997xc}.

The same $\ptmin$ and $\ptmax$ predictions are shown in Fig. \ref{fig1b},\ref{fig1d}; in this case the NNLO result is compared with the resummed result
at the NLL+NLO and NNLL+NNLO accuracy.
As expected \cite{Catani:1997xc}, resummed results do not suffer of such instabilities in the vicinity of the LO kinematical boundary; the resummed distributions are smooth and the shape is rather stable when going from NLL+NLO to NNLL+NNLO\footnote{We note that the small-$p_T$ resummation we perform in this paper is not strictly the one needed to cure these logarithmic singularities \cite{Catani:1997xc}. Nonetheless, since our resummation provides a correct description of the Higgs boson kinematics, we do not expect that a rigorous treatment of these logarithmic singularities would lead to
substantial numerical differences.}.

%Furthermore, the fixed order result is recovered out of the instability region, where the Higgs boson is produced with large transverse momentum and the resummation effects become negligible.
An analogous perturbative instability is present in the $\ptmin$ distribution around $\ptmin\sim 40$ GeV at NLO and NNLO (see Fig.~\ref{fig1a}).
Such instability, which is not related to the use of the narrow width approximation, is due to the choice of asymmetric cuts for the photons. Beyond LO, the region $\ptmin<40$ GeV opens up, and the step-like behaviour at LO leads to
integrable logarithmic singularities at NLO and beyond. The resummed NLL+NLO and NNLL+NNLO results are free of such perturbative instability.

\setcounter{footnote}{0}

Finally, a variable that is often studied
is $\cos\theta^*$, where $\theta^*$ is the polar angle of one of the photons in the Higgs boson rest frame. Given the 4-momentum of the photon $p_\gamma=(m_H/2,{\vec p_T}^{~\gamma},p_z^\gamma)$ in the Higgs rest frame, the $\theta^*$ angle is defined as follows
\begin{equation}
|\cos\theta^*|=\frac{|p_z^\gamma|}{m_H/2};
\label{eq:costheta}
\end{equation}
considering the on-shell condition for the photon $p_T^{\gamma\,2}+p_z^{\gamma\,2}=(m_H/2)^2$ and that at the LO the $p_T$ of the Higgs boson is zero, we can invert the on-shell condition, obtaining
\begin{equation}
|\cos\theta^*|=\sqrt{1-\frac{4{p}_T^{\gamma\,2}}{m_H^2}}.
\label{eq:costheta_explic}
\end{equation}

A cut on the photon transverse momentum $p_T^\gamma$ implies a maximum value for $\cos\theta^*$ at LO.
For example for $m_H=125$ GeV and $p_T^\gamma\ge 40$ GeV we obtain
\begin{equation}
p_T^\gamma\ge 40~\rm{GeV}\hspace{0.5cm} \Rightarrow \hspace{0.5cm} |\cos\theta^*|\le |\cos\theta^*_{\rm cut}|\simeq 0.768.
\label{eq:costheta_cut}
\end{equation}

\begin{figure}[!ht]
\centering
\includegraphics[width=0.495\textwidth]{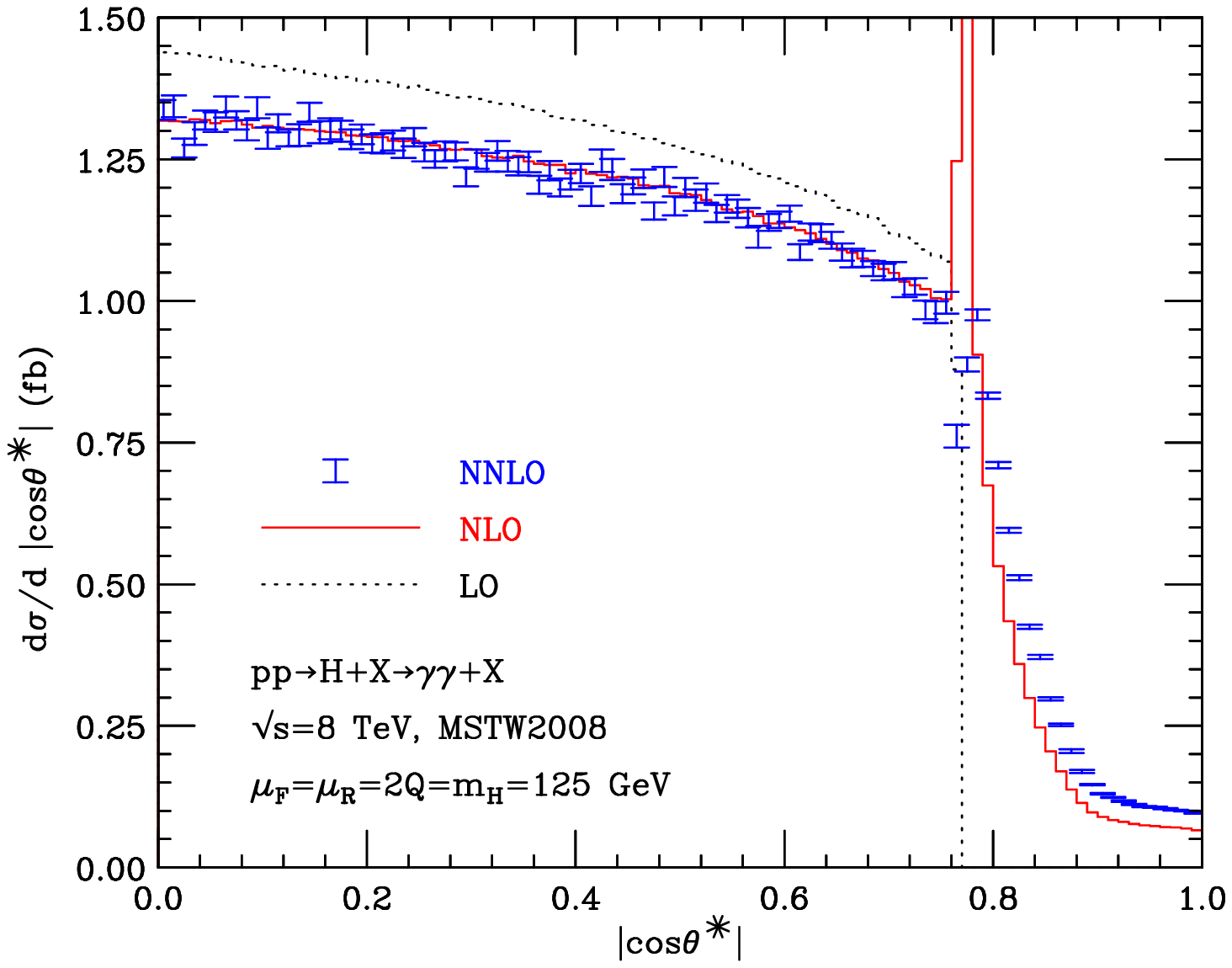}
\includegraphics[width=0.495\textwidth]{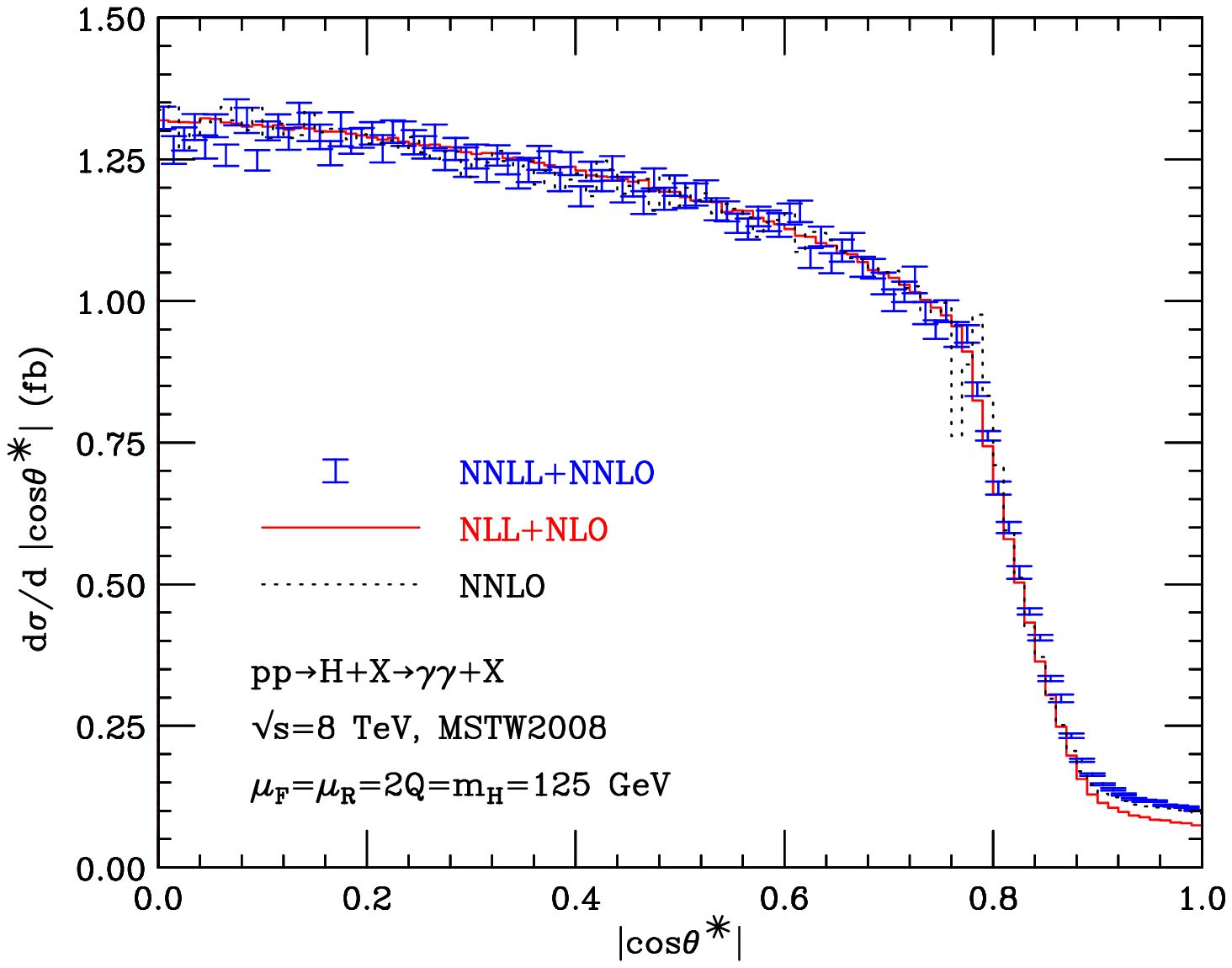}
\caption{{\em Normalised $\cos\theta^*$ distribution at the LHC. On the left: LO, NLO and NNLO results. On the right: resummed predictions at NLL+NLO and NNLL+NNLO accuracy are compared with the NNLO result.}}
\label{fig:norm_costheta}
\end{figure}

At the NLO and NNLO the Higgs transverse momentum is non vanishing and events with $|\cos\theta^*|> |\cos\theta^*_{\rm cut}|$ are kinematically allowed.
In the region of the kinematical boundary
higher-order perturbative distributions suffer of logarithmic singularities (as it happen for the photon distributions discussed above). In Fig. \ref{fig:norm_costheta} we report both the distributions (normalized to unity) obtained by fixed order and the resummed calculations.
We see that the resummed results are smooth in the region around the kinematical boundary.
Away from such region, fixed order and resummed results show perfect agreement.

\subsection{$H\to WW\to l\nu l\nu$}

We now consider the production of a SM Higgs boson with mass $m_H=140$ GeV.
The width is computed with the program HDECAY \cite{Djouadi:1997yw} to be $\Gamma_H=8.11$ MeV.
We consider the decay $W\to l\nu$ by assuming
only one final state lepton combination.
In order to isolate the possible signal some acceptance cuts are needed. Here we apply the following set of cuts \cite{Dittmaier:2012vm}:
\begin{itemize}
\item the event should contain two opposite charged leptons having $p_T$ larger than 20 GeV and in the central rapidity region $|\eta|<2.5$;
\item the missing $p_T$ of the event should be larger than 30 GeV;
\item the invariant mass of charged leptons should be larger than 12 GeV.
\end{itemize} 

In Table \ref{tab:WW}
we report the fixed order and resummed predictions for the total and accepted cross section.
We see that, also in this case, the inclusion of resummation
does not lead to substantial differences on the accepted cross section.
In Table \ref{tab:cut_WW} we report the accepted cross section for different scales. The scale uncertainty is about $\pm 17\%$ at NLL+NLO and NLO, whereas at NNLL+NNLO and NNLO it is reduced to $\pm 10\%$.

\begin{table}[!ht]
\vspace{0.3cm}
\centering
\begin{tabular}{|c||c|c|c|c|}
\hline
Cross section &\hspace{6pt} NLO\hspace{0.2cm} & NLL+NLO &\hspace{0.2cm} NNLO\hspace{0.2cm} & NNLL+NNLO\\ \hline \hline
Total [fb] & 61.58 $\pm$ 0.04 & 61.58 $\pm$ 0.04 & 76.94 $\pm$ 0.09 & 76.88 $\pm$ 0.19\\ \hline
With cuts [fb] & 20.98 $\pm$ 0.03 & 20.90 $\pm$ 0.02 & 26.44 $\pm$ 0.10  & 26.32 $\pm$ 0.05  \\ \hline
Efficiency [\%] &34.0 & 33.9 & 34.4 & 34.2\\ \hline
\end{tabular}
\caption{{\em Fixed order and resummed cross sections for $pp\rightarrow H+X\rightarrow WW+X\rightarrow l\nu l\nu+X$ before and after selection cuts.}}
\label{tab:WW}
\end{table}

\begin{table}[!ht]
\vspace{0.3cm}
\centering
\begin{tabular}{|c||c|c|c|c|}
\hline
Cross section [fb] & NLO & NLL+NLO & NNLO & NNLL+NNLO\\ \hline \hline
$(2Q=\mu_F=\mu_R)=m_H/2$ & 25.14 $\pm$ 0.03 & 25.13 $\pm$ 0.04 & 29.16 $\pm$ 0.17 & 29.05 $\pm$ 0.22 \\ \hline
$(2Q=\mu_F=\mu_R)=m_H$ & 20.98 $\pm$ 0.03 & 20.90 $\pm$ 0.02 & 26.44 $\pm$ 0.10 & 26.32 $\pm$ 0.05\\ \hline
$(2Q=\mu_F=\mu_R)=2m_H$ & 17.76 $\pm$ 0.02 & 18.26 $\pm$ 0.03 & 23.85 $\pm$ 0.07 & 24.14 $\pm$ 0.10\\ \hline
\end{tabular}
\caption{{\em Fixed order and resummed accepted cross sections for $pp\rightarrow H+X\rightarrow WW+X\rightarrow l\nu l\nu+X$ at the LHC, for different choices of the scales.}}
\label{tab:cut_WW}
\end{table}

\begin{figure}[!ht]
%\begin{center}
\centering
%\subfigure[]{\label{dec32:ptlepton1}
\includegraphics[width=0.495\textwidth]{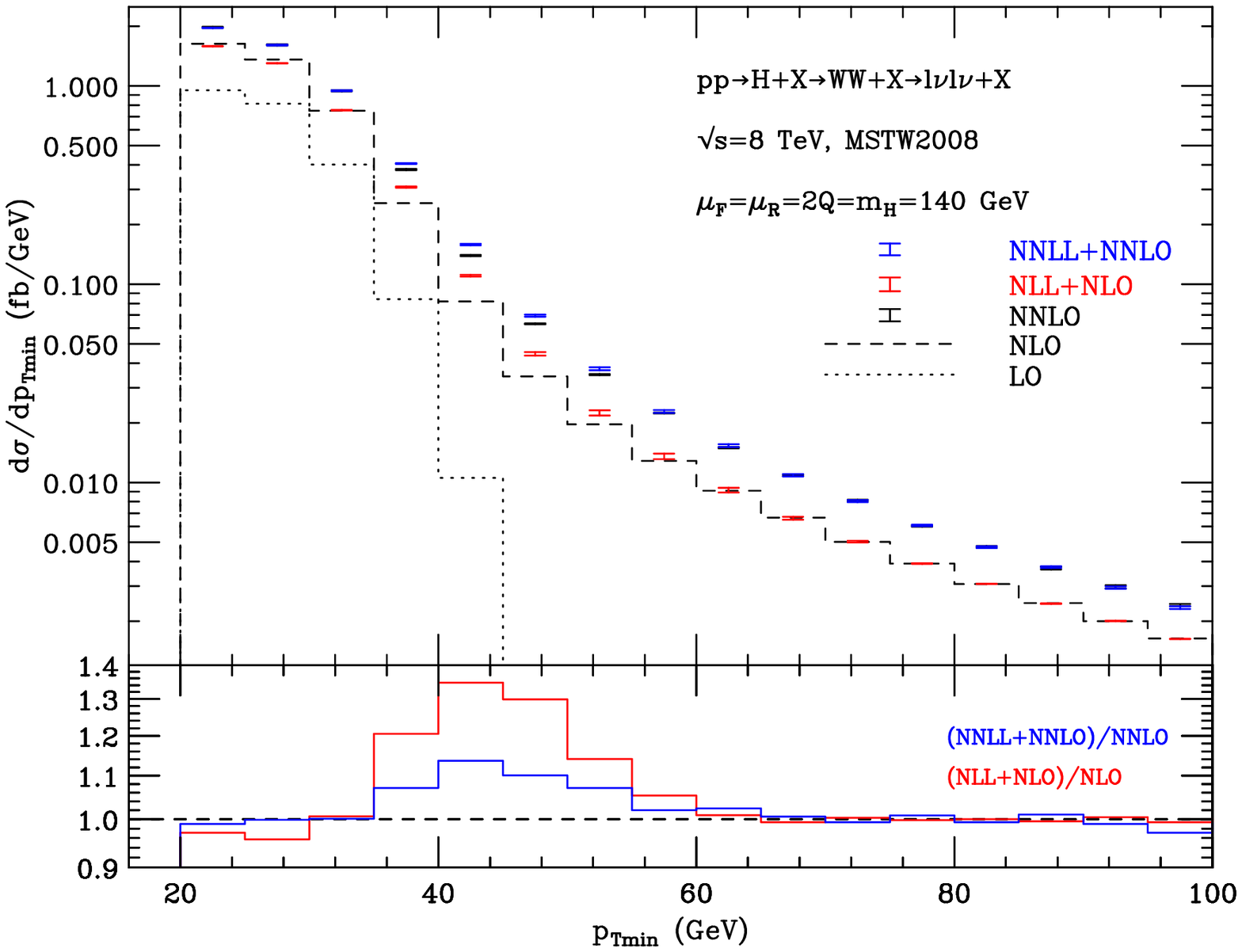}
%\subfigure[]{\label{dec32:ptlepton2}
\includegraphics[width=0.495\textwidth]{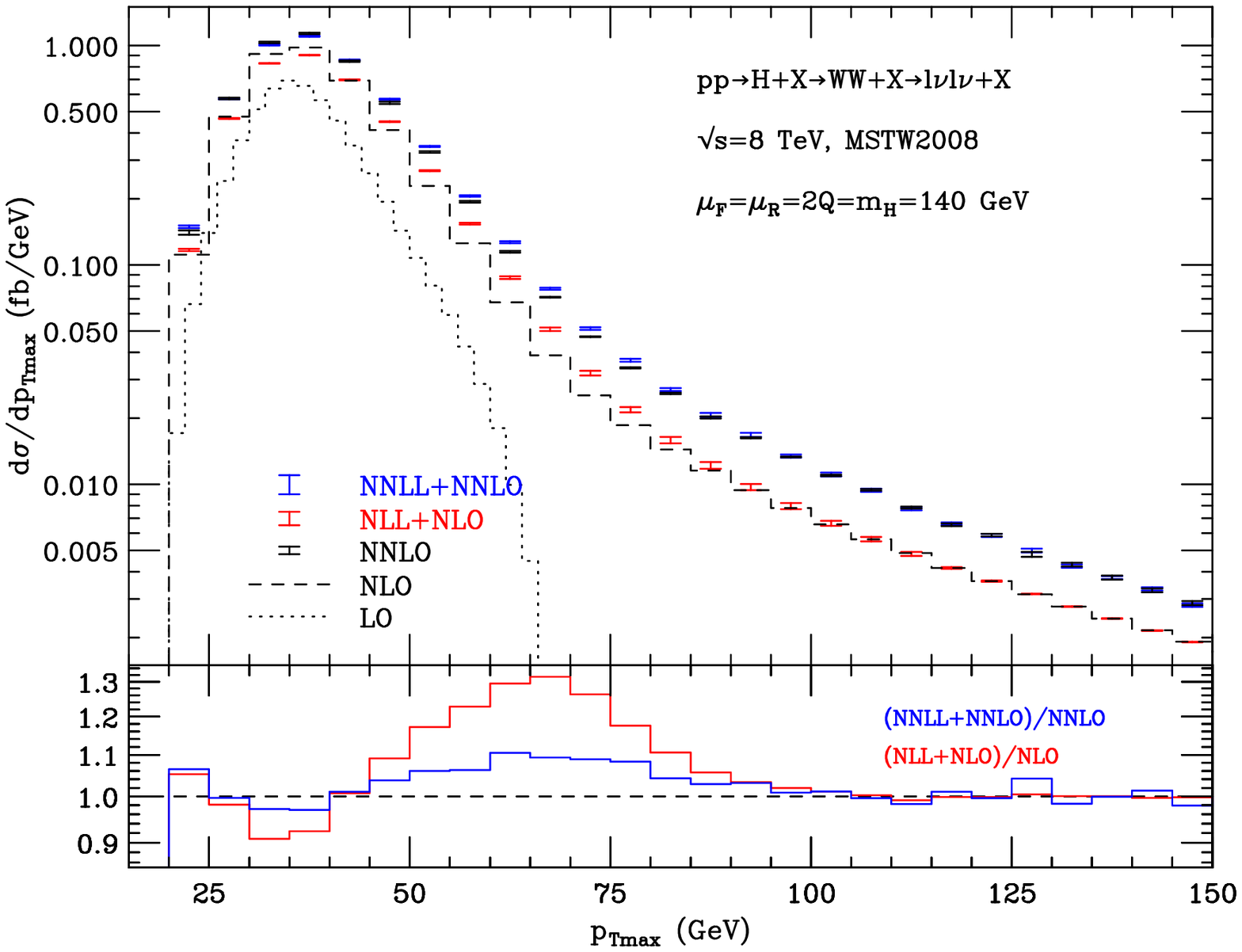}
%\vspace*{-.5cm}
%\end{center}
\caption{{\em Transverse momentum spectra of the lepton with minimum (left) and maximum (right) $p_T$ for $pp\rightarrow H+X\rightarrow WW+X\rightarrow l\nu l\nu +X$ at the LHC. Resummed results at NLL+NLO and NNLL+NNLO accuracy are compared with fixed order predictions at LO, NLO and NNLO. The lower panels show
the NNLL+NNLO result normalized to NNLO (solid) and the NLL+NLO result normalized to NLO (dashes).}}
\label{fig:pt_dec2_leptons}
\end{figure}

\begin{figure}[!ht]
\centering
\includegraphics[width=0.7\textwidth]{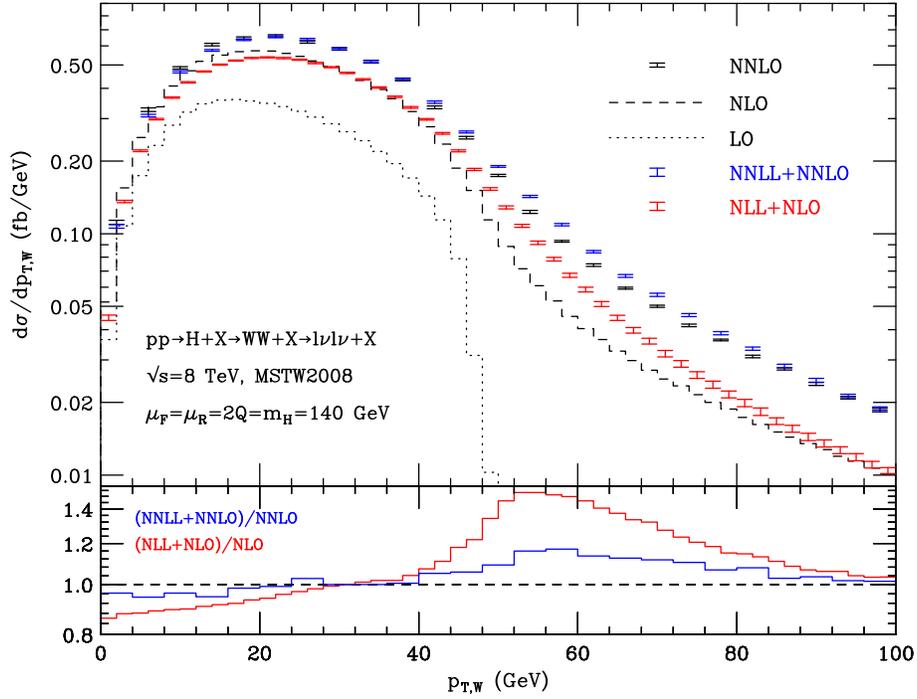}
\caption{{\em Average transverse momentum spectrum of the $W$ bosons
for $pp\rightarrow H+X\rightarrow WW+X\rightarrow l\nu l\nu +X$ at the LHC
when cuts are applied. Resummed results at NLL+NLO and NNLL+NNLO accuracy are compared with fixed order predictions at LO, NLO and NNLO. The lower panel shows
the NNLL+NNLO result normalized to NNLO (solid) and the NLL+NLO result normalized to NLO (dashes).}}
\label{fig:pt_dec2_w}
\end{figure}

For each event, we classify the transverse momenta of the charged leptons according to their minimum and maximum value (as we did for the photon transverse momenta in $H\to\gamma\gamma$). In Fig.~\ref{fig:pt_dec2_leptons} we plot the corresponding distributions. We compare the resummed NLL+NLO and NNLL+NNLO predictions with the corresponding fixed order predictions at the LO, NLO and NNLO accuracy. We see that QCD corrections tend to make the distributions harder. Analogous effects are observed on the average transverse momentum spectrum of the $W$ bosons, which is reported in Fig.~\ref{fig:pt_dec2_w}.
In particular, in order to quantitatively estimate the impact of the resummation, Figs.~\ref{fig:pt_dec2_leptons}, \ref{fig:pt_dec2_w} are organised in two panels. In the upper panels, we show the predictions obtained by different fixed order and resummed calculations. In the lower panel we plot (in red) the ratio NLL+NLO/NLO and (in blue) NNLL+NNLO/NNLO. From Fig.~\ref{fig:pt_dec2_leptons} we note that, in the peak region,
for both the $p_{Tmin}$ and $p_{Tmax}$ distributions, the resummed result is smaller by $5-10\%$ at NLL+NLO and by $2-4\%$ at NNLL+NNLO with respect to the corresponding fixed order prediction. In the intermediate region the resummation affects the results in the opposite direction, enhancing the cross section up to about $30\%$ at NLL+NLO and $10\%$ at NNLL+NNLO.  The effects observed for the average $p_T$ of the $W$ bosons (see Fig.~\ref{fig:pt_dec2_w}) are even more pronounced.
These effects on the $p_T$ spectra imply that the agreement between resummed and fixed order predictions we have observed in Table~\ref{tab:WW} cannot
persist in general. When more restrictive cuts on the transverse momenta are applied, we anticipate non negligible effects from resummation.

\begin{figure}[!ht]
\centering
\includegraphics[width=0.7\textwidth]{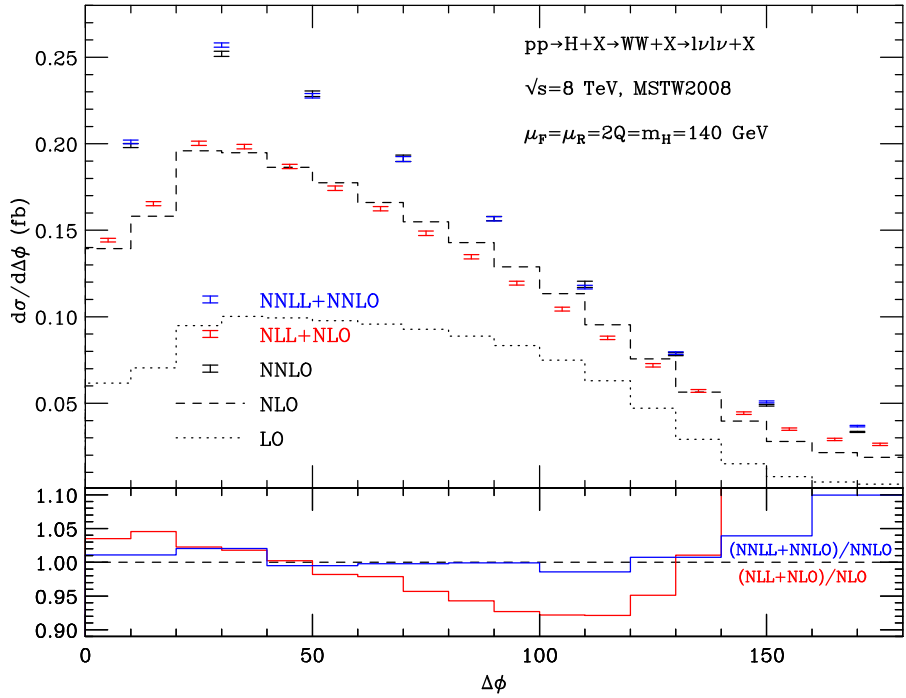}
\caption{{\em Same as in Fig.~\ref{fig:pt_dec2_w} but for the $\Delta\phi$ distribution.}}
\label{fig:dphi}
\end{figure}

A very important discriminating variable
for the $H\to WW\to l\nu l\nu$ decay channel
is the azimuthal separation of the charged leptons in
the transverse plane, $\Delta\phi$.
As is well known \cite{Dittmar:1996ss}, for the Higgs boson signal the leptons tend to be close in angle, thus the bulk of the events is produced at small $\Delta\phi$.
Our results for the $\Delta\phi$ distribution are reported in Fig. \ref{fig:dphi}. We can see that in the very small $\Delta\phi$ region ($\Delta\phi\ltap 30^o$), there are less events than expected: this is an effect of the applied cuts.
We notice that the steepness  of the distributions increases when going from LO to NLO and from NLO to NNLO, and also increases when going from fixed order to resummed predictions, i.e. from NLO to NLL+NLO and  from NNLO to NNLL+NNLO. This fact can be interpreted as follows: when the Higgs boson $p_T$ distribution is harder the final state leptons tend to be more boosted in the transverse plane and thus their transverse angular separation becomes smaller. As a consequence the steepness of the $\Delta\phi$ distribution increases and the efficiency of cuts slightly increases with the perturbative accuracy. 

\subsection{$H\to ZZ\to \mu^+\mu^-e^+e^-$}

We now consider the production of a Higgs boson with mass $m_H=150$ GeV.
The width is computed with the program HDECAY \cite{Djouadi:1997yw} to be $\Gamma_H=16.9$ MeV.
In this mass region, the $H\to ZZ\to 4l$ decay mode is not the dominant decay channel,
but still it can provide a clean and useful four lepton signature.
In the following we consider the decay of the Higgs boson in two different lepton pairs.

We consider the following cuts \cite{Dittmaier:2012vm}:
\begin{itemize}
\item the event should contain two pairs of opposite charged leptons
\item each lepton must have $p_T$ larger than 5 GeV and should be in the central rapidity region $|\eta|<2.5$;
\item for each lepton pair, the closest ($m_1$) and the next-to-closest ($m_2$) to $m_Z$ are found; then $m_1,m_2$ are required to be $m_1 > 50$ GeV and $m_2 > 12$ GeV.
\end{itemize}
Note that an isolation cut on the leptons is generally required. For example, a typical isolation
is to require the total transverse energy $E_T$ in a cone of a given radius $R$ around each lepton to be smaller than a fraction of the lepton $p_T$. As in the $H\to\gamma\gamma$ decay mode
isolation cuts cannot be applied, because in the resummed calculation we are necessarily
inclusive over the QCD radiation accompanying the Higgs boson.
By using the fixed order {\tt HNNLO} code we have checked that the numerical effect of the isolation cuts is extremely small.

In Tab.~\ref{tab:cut_ZZ} we compare the effects of cuts on the inclusive cross sections. As in the $H\to\gamma\gamma$ and $H\to WW$ decays, the efficiency slightly improves increasing the perturbative accuracy, but no substantial effects from resummation are observed. In Tab.~\ref{tab:cut_ZZ_scale} the accepted
cross section for different choices of the scales is reported.
At the NLL+NLO (NLO) accuracy the ensuing scale uncertainty is $\sim \pm 15\%$ ($\pm 17\%$) and at the NNLL+NNLO (NNLO) it is $\sim \pm 9\%$ ($\pm 10\%$).

\begin{table}[!ht]
\vspace{0.3cm}
\centering
\begin{tabular}{|c||c|c|c|c|}
\hline
Cross section &\hspace{6pt} NLO\hspace{0.2cm} & NLL+NLO &\hspace{0.2cm} NNLO\hspace{0.2cm} & NNLL+NNLO\\ \hline \hline
Total [fb] & 1.720 $\pm$ 0.001  & 1.720 $\pm$ 0.002 & 2.142 $\pm$ 0.004 & 2.156 $\pm$ 0.006  \\ \hline
With cuts [fb] & 1.127 $\pm$ 0.001 & 1.136 $\pm$ 0.001 & 1.413 $\pm$ 0.005 & 1.427 $\pm$ 0.003\\ \hline
Efficiency [\%] & 65.6 & 66.0 & 66.0 & 66.2\\ \hline
\end{tabular}
\caption{\em Fixed order and resummed cross section for $pp\rightarrow H+X\rightarrow ZZ+X\rightarrow \mu^+\mu^- e^+e^-+X$ cross section before and after geometrical acceptance cuts.}
\label{tab:cut_ZZ}
\end{table}

\begin{table}[!ht]
\vspace{0.3cm}
\centering
\begin{tabular}{|c||c|c|c|c|}
\hline
Cross section [fb] & NLO & NLL+NLO & NNLO & NNLL+NNLO\\ \hline \hline
$(2Q=\mu_F=\mu_R)=m_H/2$ & 1.350 $\pm$ 0.001 & 1.350 $\pm$ 0.004 & 1.572 $\pm$ 0.007 & 1.570 $\pm$ 0.006  \\ \hline
$(2Q=\mu_F=\mu_R)=m_H$ & 1.127 $\pm$ 0.001 & 1.136 $\pm$ 0.001 & 1.413 $\pm$ 0.005 & 1.427 $\pm$ 0.003 \\ \hline
$(2Q=\mu_F=\mu_R)=2m_H$ & 0.954 $\pm$ 0.001 & 0.992 $\pm$ 0.003 & 1.273 $\pm$ 0.003 & 1.310 $\pm$ 0.003  \\ \hline
\end{tabular}
\caption{{\em Fixed order and resummed accepted cross sections for $pp\rightarrow H+X\rightarrow ZZ+X\rightarrow \mu^+\mu^- e^+e^-+X$ at the LHC, for different choices of the scales.}}
\label{tab:cut_ZZ_scale}
\end{table}

\begin{figure}[!ht]
\begin{center}
\subfigure[]{\label{dec32:ptlepton1}
\includegraphics[width=0.485\textwidth]{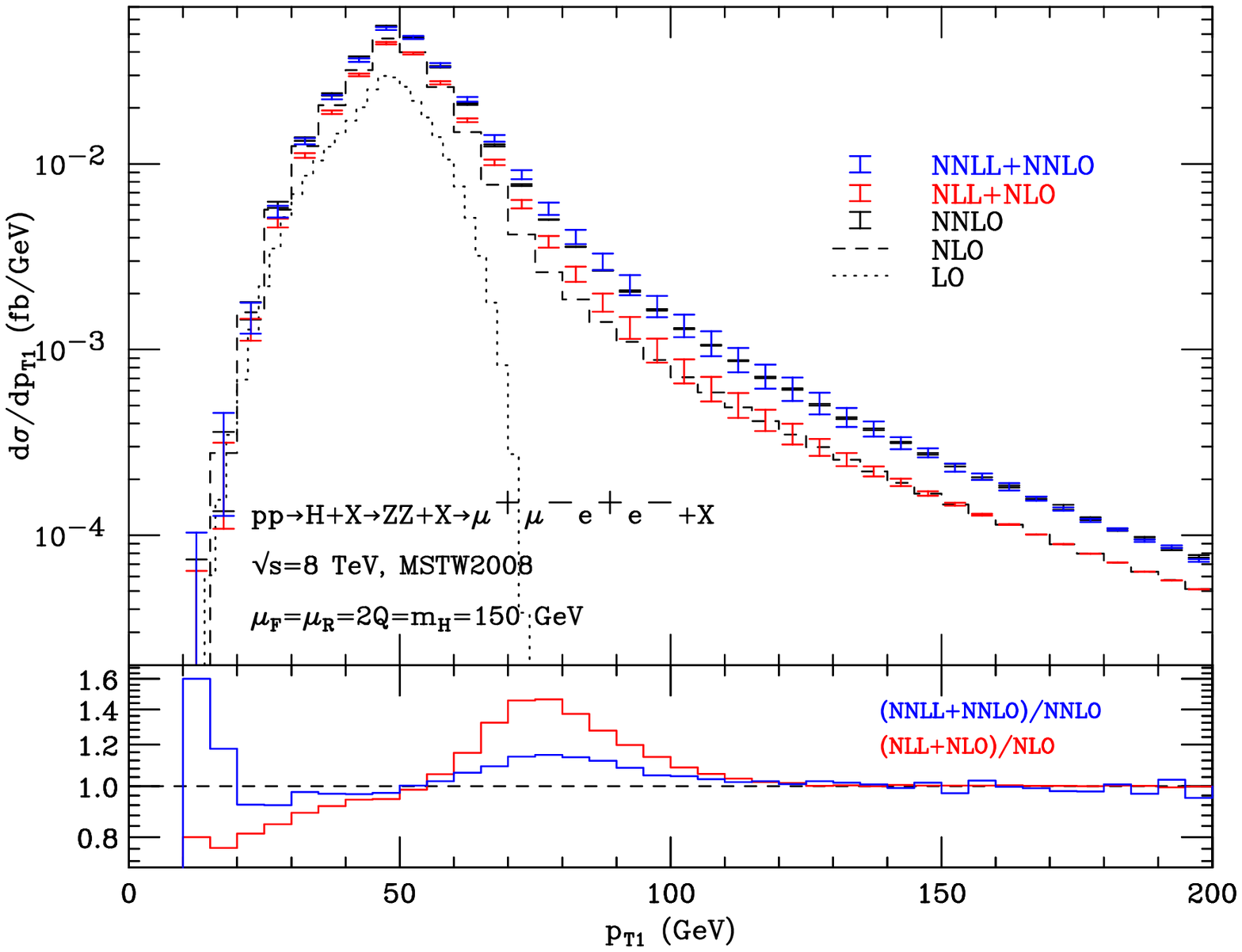}}
\subfigure[]{\label{dec32:ptlepton2}
\includegraphics[width=0.485\textwidth]{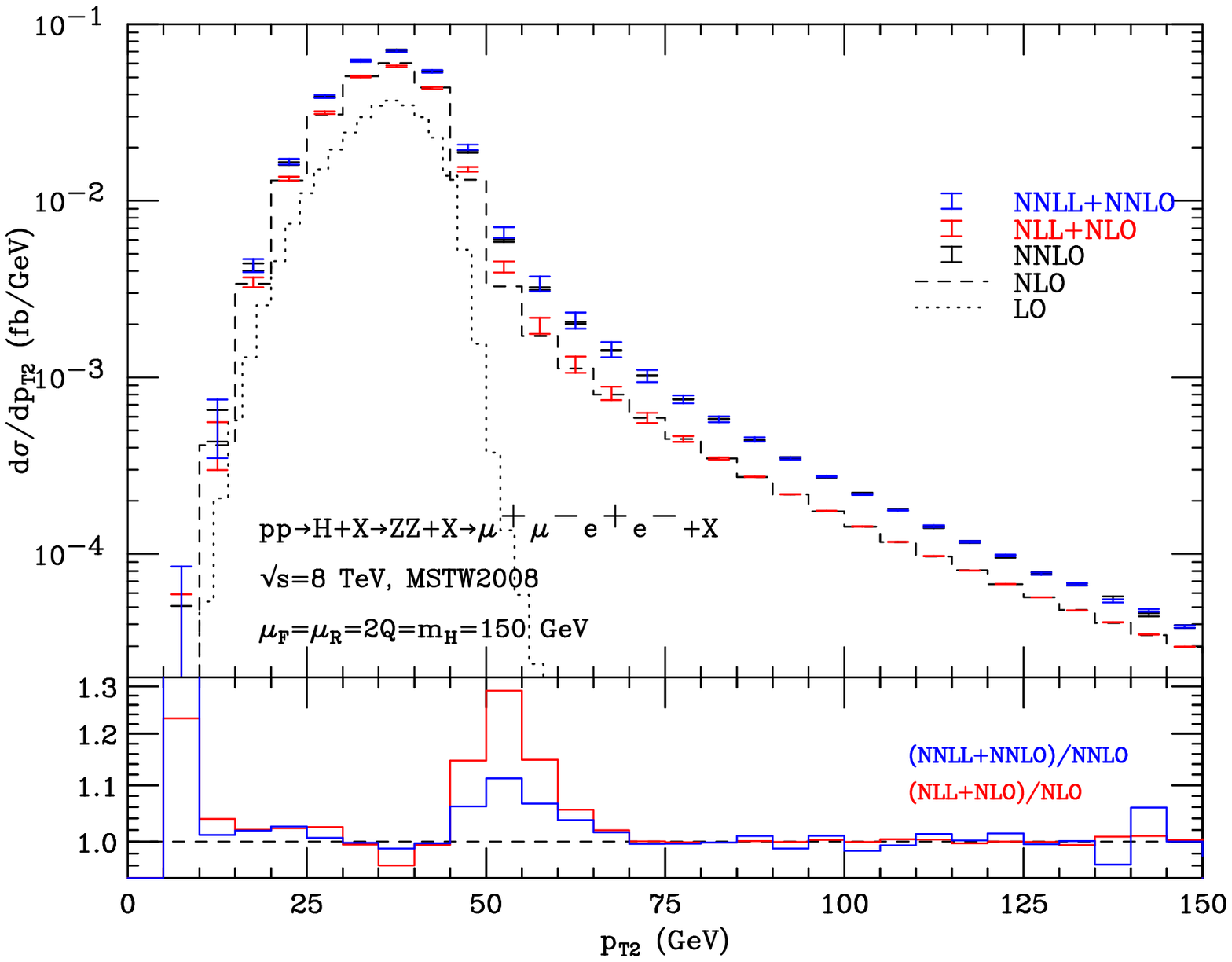}}
\subfigure[]{\label{dec32:ptlepton3}
\includegraphics[width=0.485\textwidth]{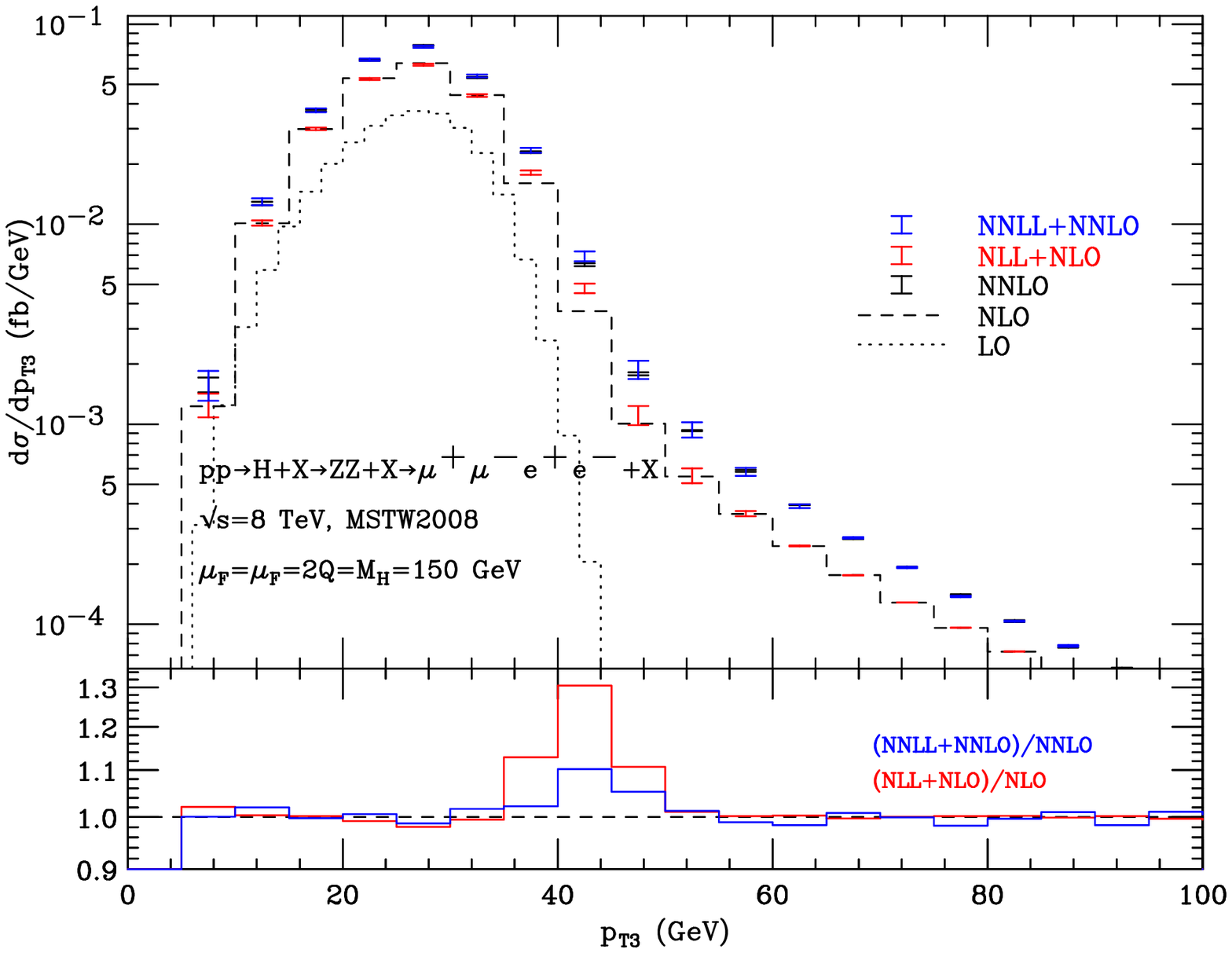}}
\subfigure[]{\label{dec32:ptlepton4}
\includegraphics[width=0.485\textwidth]{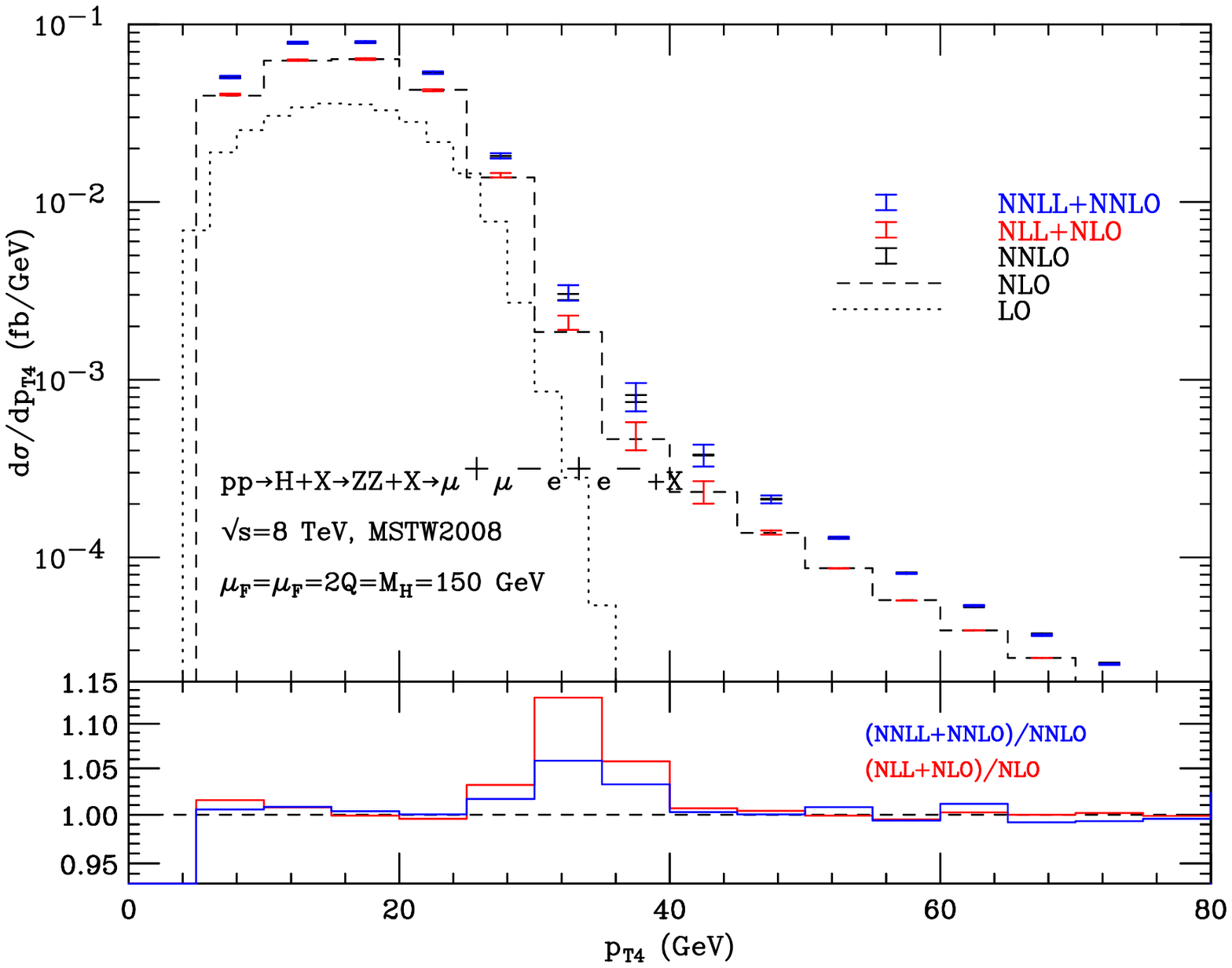}}
\vspace*{-.5cm}
\end{center}
\caption{{\em \label{dec32:ptleptons}Transverse momentum spectra of the final state leptons for $pp\rightarrow H+X\rightarrow ZZ+X\rightarrow \mu^+\mu^- e^+e^-+X$ at the LHC, when cuts are applied. The lepton $p_T$ are ordered according to decreasing $p_{T}$. They are obtained through fixed order (black) and resummed (red and blue) calculations. The lower panels show the ratios between resummed and fixed order predictions.}}
\end{figure}

In Fig. \ref{dec32:ptleptons} we plot the four $p_T$ spectra of the final state leptons. Note that at LO the $p_{T1},p_{T2}$ are kinematically bounded by $m_H/2$, whereas $p_{T3}<m_H/3$ and $p_{T4}<m_H/4$. In the vicinity of such boundaries, higher order QCD predictions may in principle develop perturbative instabilities. On the other hand, contrary to what happens in the $H\to\gamma\gamma$ decay mode, the LO distributions smoothly reach their kinematical boundary and we do not observe perturbative instabilities beyond the LO.
The impact of resummation is to make
the transverse momentum spectra harder.
The resummation effects are more pronounced in the leading lepton transverse momentum spectrum (see Fig. \ref{dec32:ptlepton1}) and less evident in the softest lepton spectrum (see Fig. \ref{dec32:ptlepton4}).

\begin{figure}[!ht]
\centering
\includegraphics[width=0.65\textwidth]{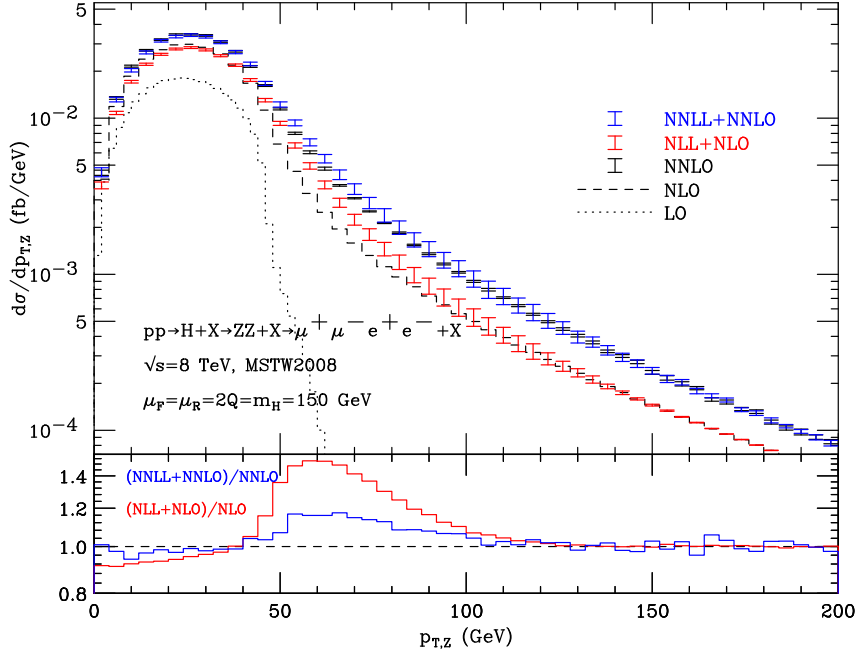}
\caption{{\em Average $p_T$ spectrum of the $Z$ bosons for $pp\rightarrow H+X\rightarrow ZZ+X\rightarrow \mu^+\mu^- e^+e^-+X$ at the LHC, when cuts are applied. Resummed results at NLL+NLO and NNLL+NNLO accuracy are compared with fixed order predictions at LO, NLO and NNLO. The lower panel shows
the NNLL+NNLO result normalized to NNLO (solid) and the NLL+NLO result normalized to NLO (dashes).}}
\label{dec32:ptZ}
\end{figure}

In Fig.~\ref{dec32:ptZ} we show the average $p_T$ distribution of the two $Z$ bosons. The comments are analogous to those for previous distributions: QCD radiation tends to make the distribution harder and the fixed order results are again recovered at large transverse momentum.

\subsection{Discussion}
\label{sec:discuss}

As explained at the end of Sec.~2 our numerical program implements
a smooth switching procedure between the resummed and fixed order results (see Eqs.~(\ref{switch}) and (\ref{fswitch})).
The numerical parameters in Eq.~(\ref{switch}) can be
consistently chosen
such that the integral of our NLL+NLO and NNLL+NNLO
resummed result still reproduces well the NLO and NNLO inclusive cross
sections.
We have studied the uncertainty inherent in such switching procedure. We find that for other forms of the switching function and different choices of the corresponding parameters the numerical effects
on the final results remain within the uncertainties estimated
by performing variations of the resummation scale $Q$.

We conclude this Section by adding few comments on the work of Ref.~\cite{Cao:2007du}. In this paper the RESBOS generator \cite{Balazs:2000wv}, which is based on the classical $b$-space resummation formalism of Ref.~\cite{Collins:1984kg},
is used to perform a study of transverse momentum resummation effects in the $H\to WW\to l\nu l\nu$ and $H\to ZZ\to 4l$ channels at the Tevatron and the LHC.
The resummed calculation in the low $p_T$ region is matched to the ${\cal O}(\as^3)$ result at high $p_T$. Besides the differences in the resummation formalism
(see Ref.~\cite{Bozzi:2005wk} for a detailed discussion) there are a few differences with respect
to the work presented here. Our calculation implements the
value of the coefficient $A^{(3)}$ from Ref.~\cite{Becher:2010tm}, whereas in Ref.~\cite{Cao:2007du}
the authors use the result of Ref.~\cite{Moch:2004pa} that applies to threshold resummation.
The calculation of Ref.~\cite{Cao:2007du}
does not include the hard collinear coefficients ${\cal H}^{(2)}$ presented
in Ref.~\cite{Catani:2011kr} and thus its accuracy, with our notations, is essentially limited to NLL+NLO (plus some of the NNLL terms).
Finally, the calculation of Ref.~\cite{Cao:2007du} does not exploit a unitarity constraint on the total cross section, and thus the normalization of the ensuing
resummed spectra is not constrained.

In their phenomenological study, when comparing resummed and fixed-order NLO predictions, the authors of Ref.~\cite{Cao:2007du} find
significant resummation effects. The reason is twofold. First, the cuts that are considered in Ref.~\cite{Cao:2007du} are more restrictive, and thus resummation effects are made more relevant. Second, the comparison is done one order
lower than ours (i.e. the NLL+NLO resummed prediction is compared to the fixed order NLO result) where we also find more significant distortions of the relevant kinematical distributions (see Figs.~\ref{fig:pt_dec2_leptons}-\ref{dec32:ptZ}).

\section{Summary and outlook}
\label{sec:summary}

We have presented a calculation of the NNLL+NNLO cross section for Higgs
boson production at the LHC, in the decay modes $H\to \gamma\gamma$,
$H\to WW\to l\nu l\nu$ and $H\to ZZ\to 4$ leptons.
The calculation takes into account some illustrative experimental cuts
analogous to the ones designed to isolate the Higgs boson signal.

Our calculation is implemented in
the numerical program {\tt HRes} \cite{grazzini_codes}.
The present version of the program includes
the most relevant decay modes of the Higgs boson, namely,
$H\to\gamma\gamma$, $H\to WW\to l\nu l\nu$ and $H\to ZZ\to 4$ leptons.
In the latter case it is possible to choose between
$H\to ZZ\to \mu^+\mu^- e^+e^-$ and $H\to ZZ\to e^+e^-e^+e^-$,
which includes the appropriate interference contribution.
The user can apply all the required cuts on
the Higgs boson and its decay products
and plot the corresponding distributions in the form of bin histograms.
These features should make our program a useful tool for Higgs searches and studies at the Tevatron and the LHC.

The calculations performed through {\tt HRes} strictly implement
the large $M_t$ approximation.
This is known to be a good approximation for the $p_T$ spectrum of the Higgs boson, provided
that $p_T$ is not too large ($p_T\ltap M_t$) \cite{Baur:1989cm}.
For very large transverse momenta the large-$M_t$ approximation is bound to fail, since the  QCD radiation accompanying the Higgs boson
becomes sensitive to the heavy-quark loop. The inclusion of top and bottom mass effects up to ${\cal O}(\as^3)$ in {\tt HRes} is feasible and is left to future work.
Another limitation of the calculation is that we completely neglect radiative corrections in the Higgs boson decay. The full QCD+EW corrections to the decay modes
$H\to WW(ZZ)\to 4$ leptons are available \cite{Bredenstein:2006rh} and we plan to include these effects
in a future version of the program.

\subsection*{Acknowledgements}
We wish to thank Stefano Catani for many helpful discussions and comments on the manuscript.
This work was supported in part by UBACYT, CONICET, ANPCyT, INFN and the Research Executive Agency (REA) of the European Union under the Grant Agree- ment number PITN-GA-2010-264564 (LHCPhenoNet). We thank the Galileo Galilei Institute for Theoretical Physics for the hospitality during the completion of this work.
%\clearpage

\end{document}